\documentstyle[aps,epsf,rotating]{revtex}

\def\bphi{\hbox{\boldmath{$\varphi$}}}
\def\bPhi{{\bf\Phi}}
\font\mb=msbm10

\begin{document}
\draft
\title{The fractality of the relaxation modes in deterministic 
reaction-diffusion systems}
\author{I. Claus and P. Gaspard\\ 
\textsl{Center for Nonlinear Phenomena 
and Complex Systems}, \textsl{Facult\'e des Sciences,}\\ 
\textsl{Universit\'e Libre de Bruxelles, Campus Plaine, Code Postal 231},\\
\textsl{B-1050 Brussels, Belgium}\\
\textsl{iclaus@ulb.ac.be}}
\maketitle
%%%%%%%%%%%%%%%%%%%%%%%%%%%%%%%%%%%%%%%%%%%%%%%%%%%%%%%%%%%%%%%%%%%%%%%%%%%%%%%
\begin{abstract}
In chaotic reaction-diffusion systems with two degrees of freedom, the modes 
governing the exponential relaxation to the thermodynamic equilibrium 
present a fractal structure which can be characterized by a Hausdorff 
dimension. For long wavelength modes, this dimension is related to 
the Lyapunov exponent and to a 
reactive diffusion coefficient. This relationship 
is tested numerically on a reactive multibaker model and on a two-dimensional 
periodic reactive Lorentz gas. The agreement with the theory is excellent.
\end{abstract}
%%%%%%%%%%%%%%%%%%%%%%%%%%%%%%%%%%%%%%%%%%%%%%%%%%%%%%%%%%%%%%%%%%%%%%%%%%%%%%%
\section{Introduction}
Recently, many studies have been devoted to fractal structures induced by the 
chaotic dynamics in the phase-space of non-equilibrium statistical systems 
and, in particular, to the fractal character of the hydrodynamic modes 
governing the exponential relaxation of the system to the thermodynamic 
equilibrium \cite
{Gilb2,Gasp3,Gilb1,Gasp1,Gasp4,Gasp5,Dorf,Gasp6,Claus2,Moran1,Evans1,Chernov1}
. This fractal character has been related to the 
entropy production in the approach of 
equilibrium, for a diffusive multibaker model \cite{Gilb2}. Previous works 
had shown similar results for the entropy production in non-equilibrium 
steady states \cite{Gasp3}. 
On the other hand, the fractal dimension characterizing the hydrodynamic modes 
of diffusion has been related to the diffusion coefficient and to the positive
Lyapunov exponent by a relation, first obtained for diffusive multibaker maps 
\cite{Gilb1}, and proved for general chaotic systems with two degrees of 
freedom \cite{Gasp1}. The construction of the hydrodynamic modes in Refs.\ 
\cite{Gilb1,Gasp1} gives us a new  
approach to relate macroscopic transport coefficients and microscopic chaotic 
properties. It establishes a link between the irreversible transport processes 
and the reversible microscopic dynamics.

Two other approaches are the so-called escape rate formalism 
and the thermostated system approach. The first approach can be applied to 
open systems with absorbing boundaries \cite{Gasp4,Gasp5,Dorf,Gasp6,Claus2}. 
The escape of trajectories leads to the formation 
of a fractal repeller, consisting in the set of orbits forever trapped within 
the absorbing boundaries. Under such conditions, the transport properties
can be related to the positive Lyapunov exponents and the Kolmogorov-Sinai 
entropy per unit time as well as to fractal dimensions 
\cite{Gasp4,Gasp5,Dorf,Gasp6,Claus2}.  
In the second approach, an external field is applied to the system, and in 
order to keep the kinetic energy constant, a special force acts on the 
particles as a heat pump \cite{Moran1,Evans1,Chernov1}. 
The system is still time-reversible but does not preserve 
phase-space volumes anymore. The trajectories converge to a fractal 
attractor. In thermostated systems, the transport properties 
have been related to the sum of Lyapunov exponents \cite{Evans1,Chernov1}.
In the new approach of Refs.\ \cite{Gilb1,Gasp1}, the advantage is that the 
fractal curve considered is directly related  to the hydrodynamic modes of 
relaxation towards thermodynamic equilibrium. It does not require absorbing 
boundaries neither a thermostat. 

In the present paper, we extend the new approach of Refs.\ \cite{Gilb1,Gasp1} 
to reaction-diffusion 
processes.  Reaction-diffusion processes are of fundamental importance to 
understand self-organization in physico-chemical systems 
\cite{Nicolis77,Nicolis95}.  In chemical systems, the microscopic mechanisms 
of the relaxation toward the thermodynamic equilibrium are still poorly 
understood already for simple reactions.  For the purpose of contributing
to this important question, we consider here a simple reactive process of 
isomerization. The particle is supposed to perform a deterministic motion 
among static scatterers and to carry  a color $A$ or $B$. 
When colliding on some special 
scatterers, playing the  role of catalysts, its color changes instantaneously 
with a given  probability $p_0$, 
\begin{eqnarray}
A + \mbox{catalytic\ scatterer}\ &\longleftrightarrow& \ B +
\mbox{catalytic\ scatterer}\ , \qquad \mbox{with\ probability} \ p_0 \ , 
\nonumber\\
A + \mbox{other\ scatterer}\ &\longleftrightarrow&\ A + \mbox{other\
scatterer} \ ,\nonumber\\ B + \mbox{other\ scatterer}\ &\longleftrightarrow& 
\ B +\mbox{other\ scatterer} \ . \label{scheme}
\end{eqnarray}
For this class of systems, we obtain an expression of the Hausdorff dimension 
of the reactive modes of relaxation in terms of the reactive dispersion 
relation  
and of a function $Q(\alpha,\beta)$ generalizing the Ruelle topological 
pressure $P(\beta)$. In the long wavelength limit, we then infer a relation 
between the Hausdorff dimension of the modes, the reactive 
diffusion coefficient, the reaction rate, and two derivatives of the function 
$Q(\alpha,\beta)$. In the limit $p_0 \to 0$, we recover a relation  previously 
derived for the diffusive case \cite{Gilb1,Gasp1}. This new expression 
relates the macroscopic transport and reaction processes to the microscopic 
underlying dynamics. 

The paper is organized as follows. In Section \ref{theory}, we generalize the 
work of Refs.\ \cite{Gasp1,Gilb1} to the reactive case considered here. 
In Section \ref{triadic},  we test our relation on a reactive multibaker 
model. 
We next study the case of a two-dimensional periodic reactive Lorentz gas in 
Section \ref{lorentz}. Conclusions are drawn in Section \ref{conclusions}.
%%%%%%%%%%%%%%%%%%%%%%%%%%%%%%%%%%%%%%%%%%%%%%%%%%%%%%%%%%%%%%%%%%%%%%%%%%%%%%%
\section{Relaxation modes: general results}
\label{theory}

We consider a system such as a Lorentz gas in which the particles move 
independently of each other along trajectories of a deterministic dynamical 
system given by
\begin{equation}
\frac{d{\bf X}}{dt}={\bf F}({\bf X})
\label{diffeq}
\end{equation}
The variables of this system are for instance the positions and velocities 
of the particle: ${\bf X}=({\bf r},{\bf v})$. The vector field 
${\bf F}({\bf X})$ is defined in this space. 
The system is spatially 
periodic in the positions $\bf r$.  The particle moves in a periodic 
lattice of scatterers given by a certain potential of interaction.

Moreover, each particle carries a color which is either $A$ or $B$.  The color 
changes when the particle interacts with some of the scatterers, called the 
catalytic scatterers according to the reaction (\ref{scheme}).  We notice that 
we may also consider that the particle has a spin one-half which flips upon 
collision on special scatterers.

We can model the reactive events by a change of color with probability $p_0$ 
when the trajectory meets a certain hypersurface $\Sigma({\bf X})=0$ defined 
in the space of the variables $\bf X$.  This hypersurface surrounds each 
catalytic scatterer and constitutes the locus where the reaction occurs.  
The introduction of the color has for consequence that the phase space is 
composed of two copies of the space of the variables $\bf X$, 
which are glued together 
along the hypersurface $\Sigma$.  This hypersurface can be viewed as the gate 
from copy corresponding to the color $A$ to the one 
corresponding to color $B$ and vice versa. These two copies now form two 
subspaces of the whole phase space of the reactive system. 
In this framework, a reactive 
event corresponds to the crossing of the hypersurface $\Sigma$ from one color 
subspace to the other.

A statistical ensemble of particles of both colors is described by two 
probability densities $q({\bf X},A,t)$ and $q({\bf X},B,t)$, to find a 
particle at the point $\bf X$ with color $c=A$ or $B$, at the 
current time $t$.  The time evolution of these probability densities is ruled 
by two coupled Liouville equations:
\begin{equation}
\cases{ \partial_t \; q({\bf X},A,t) + {\rm div}\left[{\bf F}({\bf X}) \; 
q({\bf
X},A,t)\right] = - p_0\, \Xi \left[\Sigma({\bf X})\right] \left[ q({\bf
X},A,t)-q({\bf X},B,t)\right]\ ,
\cr
\partial_t \; q({\bf X},B,t) + {\rm div}\left[{\bf F}({\bf X}) \; 
q({\bf X},B,t)
\right] = +p_0\, \Xi \left[\Sigma({\bf X})\right] \left[ q({\bf X},A,t)
-q({\bf X},B,t)\right]\ .
\cr }
\label{Liouville}
\end{equation}
where $\Xi$ is a special distribution defined as
\begin{equation}
\Xi \left[\Sigma({\bf X})\right] \ h({\bf X}) \ \equiv \ \lim_{\epsilon
\to 0^+} \ \left|{\bf F}({\bf X})\cdot \frac{\partial\Sigma}{\partial{\bf X}}
({\bf X})\right| \
\delta\left[\Sigma({\bf X})\right] \ h\left[{\bf X}-\epsilon {\bf F}({\bf X})
\right] 
\label{distrib}
\end{equation}
$\delta$ being the Dirac distribution.  The distribution (\ref{distrib}) 
expresses the fact that the particles change their color upon crossing of the 
hypersurface $\Sigma$. Accordingly, the probability densities just after the 
hypersurface $\Sigma$ in the direction of the vector field $\bf F$ are the 
following linear combinations of the densities just before $\Sigma$:
\begin{equation}
\cases{ q({\bf X}_+,A,t) = (1- p_0)\, q({\bf X}_-,A,t)+ p_0 \, 
q({\bf X}_-,B,t)\ ,
\cr
q({\bf X}_+,B,t) = p_0\, q({\bf X}_-,A,t)+ (1-p_0) \, q({\bf X}_-,B,t)\ ,
\cr}
\end{equation}
where ${\bf X}_{\pm}={\bf X}_* \pm \epsilon {\bf F}({\bf X}_*)$ with 
$\Sigma({\bf X}_*)=0$ and an arbitrarily small $\epsilon>0$.  These linear 
combinations can be expressed in terms of the transmission matrix
\begin{equation}
{\mathsf T} = \pmatrix{ 1-p_0 & p_0 \cr p_0 & 1-p_0 \cr}.
\label{transitionmatrix}
\end{equation}

We denote by ${\bf X}=\bPhi^t{\bf X}_0$ the trajectory of the differential 
equation (\ref{diffeq}) from the initial condition ${\bf X}_0$.  Moreover, we 
gather both probability densities in the vector 
\begin{equation}
{\bf q}({\bf X},t) = \pmatrix{q({\bf X},A,t)\cr q({\bf X},B,t)\cr}.
\label{qAqB}
\end{equation}
The solution of the coupled
Liouville equations (\ref{Liouville}) from the initial probability densities 
${\bf q}_0$ can then be written in the form of the following Frobenius-Perron 
operator:
\begin{equation}
{\bf q}({\bf X},t) = \left(\hat {\mathsf P}^t\cdot{\bf q}\right)({\bf X}) = 
\frac{{\mathsf T}^{N_t({\bf X})}\cdot {\bf q}_0\left( \bPhi^{-t}{\bf X}\right)}
{\left|\frac{\partial\bPhi^t}{\partial{\bf X}}(\bPhi^{-t}{\bf X})\right|}
\label{FPmatrix}
\end{equation}
with the matrix (\ref{transitionmatrix})
and where $N_t({\bf X})$ is the number of crossings through the hypersurface
$\Sigma$ performed by the segment of trajectory from
${\bf X}_0=\bPhi^{-t}{\bf X}$ to ${\bf X}=\bPhi^t{\bf X}_0$.  If the flow of 
Eq.\ (\ref{diffeq}) is volume preserving, we have that ${\rm
div}{\bf F}=0$ which implies that $\left|\frac{\partial\bPhi^t}{\partial{\bf
X}}\right|=1$, which is the case for the systems considered in the following.

The system of Liouville's equations (\ref{Liouville}) defines a random process 
and it is important to characterize the dynamical randomness generated by this 
process.  For this purpose, we calculate its Kolmogorov-Sinai entropy per unit 
time which is the amount of information produced per unit 
time by the process. The Kolmogorov-Sinai entropy is calculated by 
partitioning the whole phase space into small cells and by computing 
the probability for a trajectory to visit successively different cells of 
the partition \cite{ER}. 
If we suppose that the differential 
equation (\ref{diffeq}) defines a chaotic dynamical system with Lyapunov 
exponents $\lbrace \lambda_i\rbrace$, the dynamical instability of the
deterministic motion is a first contribution to the Kolmogorov-Sinai entropy 
given by the sum of positive 
Lyapunov exponents according to Pesin's theorem \cite{ER}. 
A second contribution comes from  
the random change of color upon crossing of the hypersurface $\Sigma$.  
At each crossing, the particle jumps from one color subspace to the other 
with probability $p_0$.  
If the number of crossings is equal to
$N_t$ during the lapse of time $t$, the frequency of $\Sigma$ crossings  
is equal to :
\begin{equation}
\nu \ \equiv \ \lim_{t\to\infty} \ \frac{\langle N_t \rangle}{t},
\label{freq}
\end{equation}
where $\langle \cdot \rangle$ denotes the average with respect to the 
equilibrium state. 
At each crossing, we have an independent dichotomous random variable with 
probabilities $(p_0,1-p_0)$, which defines a Bernoulli process with the 
rate $\nu$.  Accordingly, the overall Kolmogorov-Sinai entropy per unit 
time of the process (\ref{Liouville}) is given by
\begin{equation}
h_{\rm KS} \ = \ \sum_{\lambda_i>0} \lambda_i \ + \ \nu \ \left[ p_0 \ln
\frac{1}{p_0}+(1-p_0)
\ln \frac{1}{1-p_0}\right].
\label{KS}
\end{equation}
Pesin's expression of the Kolmogorov-Sinai entropy is recovered in the 
deterministic limits $p_0=0$ and $p_0=1$.

For $0<p_0<1$, we remark that the Kolmogorov-Sinai entropy (\ref{KS}) remains 
positive and finite as for a fully deterministic dynamical system.  The way 
the reaction is modelled here is therefore compatible with a deterministic 
dynamics that would take place near each catalytic scatterer on a time scale 
much shorter than the free flight between each scatterer.  We can imagine that 
the reaction follows a complicated trajectory on such a short time scale in a 
complicated potential, which is here simplified by capturing the complexity of 
the reaction into a sole transition probability $p_0$.  Still the so-defined
random process has a finite Kolmogorov-Sinai entropy contrary to stochastic 
processes which have an infinite Kolmogorov-Sinai entropy per unit time.  
Indeed, it has been shown elsewhere
\cite{GaspWang,Gaspbook} that stochastic processes such as the birth-and-death 
processes or the linearized Boltzmann processes have a dynamical randomness on 
arbitrarily small scales $\epsilon$ with an $\epsilon$-entropy per unit time 
increasing as $h(\epsilon)\sim\ln(1/\epsilon)$.  Stochastic processes defined 
by Langevin stochastic differential equations have an even larger dynamical 
randomness with an $\epsilon$-entropy per unit time increasing as 
$h(\epsilon)\sim1/\epsilon^2$.  The Kolmogorov-Sinai entropy is the supremum 
of the $\epsilon$-entropy in the limit $\epsilon\to 0$ and is therefore 
infinite.  In contrast with such stochastic processes, the present random 
process has a finite Kolmogorov-Sinai entropy.  This dynamical randomness 
is comparable with the randomness of the classical dynamics on graphs which 
has been recently characterized \cite{BarraGasp}.  For the classical graphs, 
the motion is one-dimensional so that the only source of randomness is the 
random bifurcations at the vertices of the graph.  Here, there is an extra 
source of randomness which is the dynamical instability due to the
chaotic motion of the particle in the multi-dimensional phase space of 
Eq.\ (\ref{diffeq}).

The overall dynamical randomness (\ref{KS}) induces a relaxation of the 
probability densities 
toward the thermodynamic equilibrium.  In order to study this relaxation we 
have to find the relaxation rates which are given by the eigenvalues of the 
Frobenius-Perron operator (\ref{FPmatrix}).  It turns out that the Liouville 
equations (\ref{Liouville}) can be decoupled if we introduce the probability 
density $f$ to find a particle at the point $\bf X$ no matter which is its 
color, and the difference $g$ between both color densities:
\begin{eqnarray}
f({\bf X},t)  &=&  q({\bf X},A,t) +  q({\bf X},B,t) \\
g({\bf X},t)  &=&  q({\bf X},A,t) -  q({\bf X},B,t) .
\end{eqnarray}
These new functions are gathered in the vector
\begin{equation}
{\bf q}_{\rm d}({\bf X},t) = \pmatrix{f({\bf X},t)\cr g({\bf X},t)\cr}
\end{equation}
which is related to the previous vector (\ref{qAqB}) by the relation
\begin{equation}
{\bf q}_{\rm d} = {\mathsf S}\cdot {\bf q}
\end{equation}
in terms of the matrix
\begin{equation}
{\mathsf S}= \pmatrix{ 1 & 1 \cr 1 & -1 \cr}.
\end{equation}
This matrix diagonalizes the transition matrix (\ref{transitionmatrix}) into
\begin{equation}
{\mathsf T}_{\rm d}= {\mathsf S}\cdot{\mathsf T}\cdot{\mathsf S}^{-1}=
\pmatrix{ 1 & 0 \cr 0 & 1-2p_0 \cr}.
\end{equation}
Accordingly, the Liouville equations (\ref{Liouville}) decouple into
\begin{equation}
\cases{ \partial_t \; f({\bf X},t) + {\rm div}\left[{\bf F}({\bf X}) \; f({\bf
X},t)\right] = 0\ ,
\cr
\partial_t \; g({\bf X},t) + {\rm div}\left[{\bf F}({\bf X}) \; g({\bf X},t)
\right] = -2p_0\, \Xi \left[\Sigma({\bf X})\right] g({\bf X},t)\ .
\cr}
\label{newLiouville}
\end{equation}
which shows that the probability density $f$ obeys the deterministic 
Liouville equation while the difference $g$ solely controls the reaction.  
Similarly, the Frobenius-Perron operator (\ref{FPmatrix}) splits in the 
two following operators:
\begin{eqnarray}
f({\bf X},t) &=& \frac{f_0\left( \bPhi^{-t}{\bf X}\right)}{\left|
\frac{\partial\bPhi^t}{\partial{\bf X}}(\bPhi^{-t}{\bf X})\right|} 
\label{opdiff}\\
g({\bf X},t) &=& (1-2p_0)^{N_t({\bf X})}\frac{
g_0\left( \bPhi^{-t}{\bf X}\right)}{\left|
\frac{\partial\bPhi^t}{\partial{\bf X}}(\bPhi^{-t}{\bf X})\right|}. 
\label{opreac}
\end{eqnarray}

An eigenvalue problem can now be considered separately for both operators 
(\ref{opdiff}) and (\ref{opreac}).  If we denote them generically by 
$\hat P^t$, we are looking for eigenstates given by some distributions which 
are quasiperiodic in the position space $\bf r$ with a wavenumber $\bf k$:
\begin{equation}
\Psi_{\bf k}({\bf X}) =  e^{i{\bf k}\cdot {\bf r}({\bf X})} \ 
\Upsilon_{\bf k}({\bf X})
\label{eigen}
\end{equation}
where $\Upsilon_{\bf k}$ is periodic in $\bf r$.  The wavenumber $\bf k$ 
describes spatial inhomogeneities of wavenumber $\ell=2\pi/\Vert{\bf k}\Vert$.  
The distribution (\ref{eigen}) is supposed to be a
solution of the eigenvalue problem
\begin{equation}
\hat P^t \Psi_{\bf k} = e^{s_{\bf k}t} \Psi_{\bf k}
\end{equation}
associated with the operator $\hat P^t$.   
The quantity $-s_{\bf k}$ is the relaxation rate of the eigenstate 
(\ref{eigen}).  

The idea is that these eigenstates can be obtained in principle by applying the
operator $\hat P^t$ over a long time to some expression (\ref{eigen}) 
and by renormalizing the result in order to compensate the relaxation 
according to
\begin{equation}
\Psi_{\bf k}({\bf X}) = \lim_{t\to\infty} e^{-s_{\bf k}t} \ \hat P^t 
\left[ e^{i{\bf k}\cdot {\bf r}({\bf X})} \ \Upsilon({\bf X})\right]
\label{limit}
\end{equation}
even starting from some arbitrary function $\Upsilon$.  This
renormalization-semigroup idea has been developed in detail for diffusion in 
Refs.\ \cite{Gaspbook,Eigen}.  For chaotic dynamical systems (\ref{diffeq}), 
such eigenstates turn out to be singular distributions without a density 
function but with cumulative functions defined by
\begin{equation}
F_{\bf k}(\theta) = \int_0^{\theta} d\theta' \Psi_{\bf k}\left[{\bf X}
(\theta')\right]
\end{equation}
where $\theta$ is a parameter (such as an angle) of a curve ${\bf X}(\theta)$ 
defined in the phase space.

The hydrodynamic modes of diffusion as well as the reactive modes can be 
identified as the eigenstates corresponding to the slowest relaxation rates 
and they can be constructed thanks to the above idea of renormalization 
semigroup, as we shall show below.

\subsection{Diffusive modes}

In order to construct a diffusive mode of wavenumber ${\bf k}$ we consider the
Frobenius-Perron operator (\ref{opdiff}) and an initial distribution 
$\Upsilon({\bf X})$ of particles of initial positions ${\bf r}_0$.  We 
integrate the trajectories by using the equations of motion (\ref{diffeq}) in 
order to obtain the position ${\bf r}_t$ after a long time $t$.  
As a consequence of the aforementioned idea of renormalization semigroup,
the relaxation rate of the diffusive mode is given by the decay rate of the 
Van Hove intermediate incoherent scattering function \cite{VanHove,Boon} as 
\begin{equation}
s_{{\bf k}}=\lim_{t \to \infty}\ \frac{1}{t}\ \ln \langle \exp[i{\bf k}\cdot
({\bf r}_t-{\bf r}_0)]\rangle \label{VanHove}
\end{equation}
where $\langle \cdot\rangle$ is an average over initial conditions distributed 
according to some distribution $\Upsilon$ \cite{Gasp1,Gaspbook,Eigen}.  
In a similar way, the cumulative function of the diffusive mode 
can be obtained as
\begin{equation}
F_{\bf k}(\theta)\equiv\lim_{t \to \infty} \frac{\int_0^{\theta}d\theta'\ 
\exp\{i{\bf k}\cdot[{\bf r}_t(\theta')-{\bf r}_0(\theta')]\}}
{\int_0^{2\pi}d\theta'\ \exp\{i{\bf k}\cdot[{\bf r}_t(\theta')-
{\bf r}_0(\theta')]\}}
\label{cumul}
\end{equation}
where the initial conditions are distributed according to a distribution 
$\Upsilon$ which is here taken as a uniform distribution along a 
one-dimensional line parametrized by the angle 
$\theta \; \in  [0,2\pi[$ \cite{Gasp1}.  

For two-degree-of-freedom chaotic systems with one positive 
Lyapunov exponent $\lambda$, the cumulative function (\ref{cumul}) forms a 
fractal curve in the complex plane and its Hausdorff dimension is given in 
terms of the diffusion coefficient and the Lyapunov exponent according to
\begin{equation}
D_{\rm H}({\bf k})=1+\frac{\cal D}{\lambda}k^2+{\mathcal O}(k^4) 
\label{diff}
\end{equation} 
as shown in Refs.\ \cite{Gilb1,Gasp1}.  The derivation of Ref.\ \cite{Gasp1} is
here summarized because we need it for the generalization to the reactive 
modes. 

We assume that we can use the result by Sinai, Bowen 
and Ruelle that averages can be performed in terms of unstable trajectories 
covering the phase space within a certain resolution 
\cite{Sinai,Bowen,Ruelle}. 
The probability weight given to each of these trajectories is inversely 
proportional to their stretching factor $\Lambda_t^{(i)}$ which characterizes 
their dynamical instability. Accordingly, Eq.\ (\ref{VanHove}) can be 
transformed  into the condition
\begin{equation}
\sum_j |\Lambda_t^{(j)}|^{-1}\ \exp(-s_{{\bf k}}t)\ \exp\{i{\bf k}\cdot
[{\bf r}_t^{(j)}-{\bf r}_0^{(j)}]\}\sim_{t \to \infty}\ 1 \ .
\label{ratesum}
\end{equation}
If the sum in Eq.\ (\ref{ratesum}) is restricted to the trajectories issued 
from initial conditions in the interval $[0,\theta]$, we obtain at time $t$ 
a polygonal approximation of the cumulative function (\ref{cumul}): 
the integral $\int_0^{\theta}$ is an average over trajectories with initial 
conditions in $[0,\theta]$ and the denominator is proportional to the factor 
$\exp (s_{\bf k}t)$. 
At time $t$, the curve $({\rm Re}\; F_{\bf k}, {\rm Im}\; F_{\bf k})
\subset \hbox {\mb C}$ is thus
approximated by a polygon of sides given by the small complex vectors
\begin{equation}
\Delta F^{(j)} = \vert\Lambda_t^{(j)}\vert^{-1} \; \exp(-s_{\bf k}t) \;
\exp\left\{ i{\bf k}\cdot\left[{\bf
r}_t^{(j)}-{\bf r}_0^{(j)}\right]\right\} \;.
\end{equation}
Each side has the length 
\begin{equation}
\varepsilon_j = \vert\Delta F^{(j)}\vert = \vert\Lambda_t^{(j)}\vert^{-1}
\; \exp(-{\rm Re}\; s_{\bf k}\; t)
\; .
\end{equation}
In the limit $t \to \infty$, this polygon converges to a fractal curve, 
characterized by a Hausdorff dimension given by 
$\sum_j \varepsilon_j^{D_{\rm H}}\sim 1$. Accordingly, the Hausdorff 
dimension of the hydrodynamic mode should satisfy the condition
\begin{equation}
\sum_j \vert\Lambda_t^{(j)}\vert^{-D_{\rm H}} \; \exp(-D_{\rm H}\; {\rm
Re}\; s_{\bf k}\; t) \sim_{t\to\infty} 1
\; .
\end{equation}

On the other hand, Ruelle's topological pressure is defined in dynamical 
systems theory by
\begin{equation}
P(\beta)\equiv \lim_{t \to \infty}\; \frac{1}{t}\;\ln\langle 
|\Lambda_t|^{1-\beta}  \rangle
\label{press}
\end{equation}
where the average $\langle \cdot\rangle$ is carried out over the equilibrium 
invariant measure \cite{Ruelle}. The mean positive Lyapunov exponent of the 
system is given by $\lambda=-dP/d\beta |_{\beta=1}$ and
in this system without escape, $P(1)=0$.
Equation (\ref{press}) can be transformed into
\begin{equation}
\sum_j \vert\Lambda_t^{(j)}\vert^{-\beta} \; \exp\left[-P(\beta)\; t\right]
\sim_{t\to\infty} 1 \; \label{press2}.
\end{equation}
Comparing with Eq.\ (\ref{ratesum}), we obtain 
\begin{equation}
P(D_{\rm H})=D_{\rm H}\ {\rm Re}\ s_{\bf k}\; .\label{result}
\end{equation}
If the wavenumber ${\bf k}$ vanishes, the cumulative function (\ref{cumul})
becomes $F_{{\bf k}=0}(\theta)=\theta/2\pi$, which forms a straight 
line in the complex plane. In this equilibrium limit, the relaxation rate 
vanishes, $s_{{\bf k}=0}=0$, so that $P(D_{\rm H})=0$. It means for this 
system without escape that $D_{\rm H}=1$, as it should be for a straight 
line. For a non-vanishing but small wavenumber, the Hausdorff dimension 
is expected to deviate from unity. Inserting $D_{\rm H}=1+\delta$ and the 
dispersion relation $s_{\bf k}=-{\cal D}k^2+{\mathcal O}(k^4)$ in 
Eq.\ (\ref{result}), we can expand both sides in powers of the wavenumber, 
using the aforementioned properties of the topological pressure. 
This straightforward calculation shows that the Hausdorff dimension of 
the hydrodynamic mode is given by Eq.\ (\ref{diff}) \cite{Gasp1}.
%%%%%%%%%%%%%%%%%%%%%%%%%%%%%%%%%%%%%%%%%%%%%%%%%%%%%%%%%%%%%%%%%%%%%%%%%%%%%%

\subsection{Reactive modes}
\label{reacmodes}
\subsubsection{General results}
We are now going to extend this result to the reactive modes of the 
isomerization process (\ref{scheme}).  The modes of interest here are 
the ones describing the evolution of the difference between the two colors 
$A$ and $B$, a quantity which relax to zero when the system reaches 
equilibrium. In this case, we consider the operator (\ref{opreac}) which 
differs from the diffusive Frobenius-Perron operator (\ref{opdiff})
by the extra factor $(1-2p_0)^{N_t}$.  Accordingly, this factor should 
multiply the exponential $\exp(i{\bf k}\cdot{\bf r})$ in the expression 
(\ref{VanHove}) in order to obtain the relaxation rate of the reactive modes, 
which is thus given by
\begin{equation}
s_{{\bf k}}=\lim_{t \to \infty}\ \frac{1}{t}\ \ln \langle (1-2p_0)^{N_t} 
\exp[i{\bf k}\cdot({\bf r}_t-{\bf r}_0)]\rangle \label{VanHovereac}
\end{equation}
where $N_t$ is the number of collisions performed on catalysts during the 
time $t$ \cite{Claus1}.
The cumulative function of the reactive modes are similarly given by
\cite{Claus1}
\begin{equation}
F_{{\bf k}}(\theta)\equiv\lim_{t \to \infty} \frac{\int_0^{\theta}d\theta'\ 
(1-2p_0)^{N_t(\theta')} \exp\{i{\bf k}\cdot[{\bf r}_t(\theta')-{\bf r}_0
(\theta')]\}}{\int_0^{2\pi}d\theta'\ (1-2p_0)^{N_t(\theta')} 
\exp\{i{\bf k}\cdot[{\bf r}_t(\theta')-{\bf r}_0(\theta')]\}} \; .
\label{cumulreac}
\end{equation}
We notice that Eqs.\ (\ref{VanHovereac}) and (\ref{cumulreac}) consistently 
reduce to Eqs.\ (\ref{VanHove}) and (\ref{cumul}) in the purely diffusive 
limit $p_0\to 0$. 

As for the diffusive case, Eq.\ (\ref{VanHovereac}) can be transformed into 
the condition
\begin{equation}
\sum_j |\Lambda_t^{(j)}|^{-1}\ (1-2p_0)^{N_t^{(j)}} \exp(-s_{{\bf k}}t)\ 
\exp\{i{\bf k}\cdot[{\bf r}_t^{(j)}-{\bf r}_0^{(j)}]\}\sim_{t \to \infty}\ 1 
\; .
\label{ratesumreac}
\end{equation}
The curve $({\rm Re}\; F_{\bf k}, {\rm Im}\; F_{\bf k})
\subset \hbox {\mb C}$ is approximated at time $t$ by a polygon of sides 
given by
\begin{equation}
\varepsilon_j = \vert\Delta F^{(j)}\vert = \vert\Lambda_t^{(j)}\vert^{-1}
\; \vert 1-2p_0\vert^{N_t^{(j)}} \;\exp(-{\rm Re}\; s_{\bf k}\; t)\; .
\end{equation}
In the limit $t \to \infty$, this polygon converges to a fractal curve, 
characterized by a Hausdorff dimension given by 
$\sum_j \varepsilon_j^{D_{\rm H}}
\sim 1$. Accordingly, the Hausdorff dimension of the reactive mode 
should satisfy the condition
\begin{equation}
\sum_j \vert\Lambda_t^{(j)}\vert^{-D_{\rm H}} \; \vert 1-2p_0\vert^{N_t^{(j)}
D_{\rm H}}\;\exp(-D_{\rm H}\; {\rm
Re}\; s_{\bf k}\; t) \sim_{t\to\infty} 1
\; .
\end{equation}
By analogy with Eq.\ (\ref{press2}), we can define a function 
$Q(\alpha,\beta)$ by
\begin{equation}
\sum_j \vert\Lambda_t^{(j)}\vert^{-\beta} \; \vert 1-2p_0 \vert^{
\alpha N_t^{(j)}}\;
\exp\left[-Q(\alpha,\beta)t\right]\; 
\sim_{t\to\infty} 1 \; \label{Q2}
\end{equation}
which can be rewritten as 
\begin{equation}
Q(\alpha,\beta)=\lim_{t \to \infty}\frac{1}{t}\;\ln \langle (1-2p_0)^
{\alpha N_t} \; \vert \Lambda_t  \vert^{1-\beta} \rangle \; . \label{Qdef}
\end{equation}
Comparing Eq.\ (\ref{Q2}) with Eq.\ (\ref{ratesumreac}), we obtain 
\begin{equation}
Q(D_{\rm H},D_{\rm H})=D_{\rm H}\ {\rm Re}\ s_{\bf k} \; .\label{resultreac}
\end{equation}
The relation between this new function and the topological pressure is given 
by
\begin{equation}
P(\beta)=Q(0,\beta) \; .
\end{equation} 
If the wavenumber ${\bf k}$ vanishes, the cumulative function 
(\ref{cumulreac}) is reduced to
\begin{equation}
F_{{\bf k}=0}(\theta)=\lim_{t \to \infty}\frac{\int_0^{\theta}d\theta' 
(1-2p_0)^{N_t(\theta')}}{\int_0^{2\pi}d\theta' (1-2p_0)^{N_t(\theta')}} \; .
\label{cum0reac}
\end{equation}
This is not a straight line as in the diffusive case: each initial condition 
is here weighted by a factor $(1-2p_0)^{N_t}$. Provided that 
$p_0<0.5$, this means that the 
contribution of a given initial condition to $F_0$ will be smaller if the 
corresponding trajectory meets many catalysts, and vice versa. 
But the Hausdorff dimension of $F_0$ is anyway equal to unity. Indeed, 
for ${\bf k}=0$, the relaxation rate is given by 
\begin{equation}
s_{{\bf k}=0}=\lim_{t \to \infty}\frac{1}{t}\;\ln\;\langle 
(1-2p_0)^{N_t} \rangle ,\label{s0}
\end{equation} 
which is also equal to $Q(1,1)$ according to  the 
definition (\ref{Qdef}). Using these equalities, we can deduce from 
(\ref{resultreac}) that $D_{\rm H}=1$ for vanishing wavenumber. 
Again the Hausdorff dimension will deviate from unity for a non-vanishing 
wavenumber. The reactive dispersion relation is given by $s_{\bf k}=-2\kappa
-{\cal D}^{\rm (r)}k^2+{\mathcal O}(k^4)$. If we compare it 
with (\ref{s0}) at vanishing wavenumber, we obtain 
\begin{equation}
Q(1,1)=-2\kappa=\lim_{t \to \infty}\;\frac{1}{t}\;\ln\langle (1-2p_0)^{N_t} 
\rangle \; . \label{reacrate}
\end{equation}
Using this equality, the dispersion relation and the expression 
$D_{\rm H}(k)=1+ak^2+{\mathcal O}(k^4)$, Eq.\ (\ref{resultreac}) can be 
expanded in powers of the wavenumber, giving  an expression of the Hausdorff 
dimension of the reactive mode
\begin{equation}
D_{\rm H}({\bf k})=1-\frac{{\cal D}^{\rm (r)}k^2}{2\kappa+\frac{\partial Q}
{\partial \alpha}(1,1)+\frac{\partial Q}{\partial \beta}(1,1)}
+{\mathcal O}(k^4) \; .\label{hausdim}
\end{equation}

An expression for the two partial derivatives of $Q(\alpha, \beta)$ can be 
directly obtained from Eq.\ (\ref{Qdef}), provided that $p_0 < 0.5$:
\begin{equation}
\frac{\partial Q}{\partial \alpha}(1,1)= \ln(1-2p_0) \ \lim_{t \to \infty} 
\frac{\langle N_t (1-2p_0)^{N_t}  \rangle}{t\; \langle (1-2p_0)^{N_t}  
\rangle}
\end{equation}
and
\begin{equation}
\frac{\partial Q}{\partial \beta}(1,1)=- \lim_{t \to \infty}
\frac{\langle (1-2p_0)^{N_t} \ln|\Lambda_t|  \rangle}{t \; \langle 
(1-2p_0)^{N_t}  \rangle}\; .
\end{equation}

Similarly, the reactive diffusion coefficient is given by
\begin{equation}
{\cal D}^{\rm (r)} = \lim_{t\to\infty} \frac{\langle (1-2p_0)^{N_t}\; 
({\bf r}_t-{\bf r}_0)^2\rangle}{2 \; d \; t \; \langle (1-2p_0)^{N_t}\rangle}
\label{reacdiffcoeff}
\end{equation}
where $d$ is the dimension of the space of positions $\bf r$ where diffusion 
occurs.

\subsubsection{Limit of small reaction probability $p_0 \ll 1$}

Let us now consider the limit case $p_0 \ll 1$. The three terms of the 
denominator appearing in Eq.\ (\ref{hausdim}) can be expanded in powers 
of $p_0$.  The reaction rate is given by
\begin{eqnarray}
2 \kappa &=& \lim_{t \to \infty}\; \frac{1}{t} \left[ \;2\; p_0\; 
\langle N_t
\rangle  + 2\; p_0^2\;\langle N_t \rangle+\frac{8}{3}\;p_0^3\;\langle N_t 
\rangle
\right.\nonumber\\
& &-2\; p_0^2\; \left(\langle N_t^2 \rangle-\langle N_t \rangle^2 \right)
-4\;p_0^3\; \left(\langle  N_t^2\rangle-\langle N_t \rangle^2 \right)
\nonumber\\
& & \left. +4\;p_0^3\; \left(\frac{1}{3}\;\langle N_t^3 \rangle
+\frac{2}{3}\;\langle N_t \rangle^3-\langle N_t \rangle\;\langle N_t^2 
\rangle \right)
+{\mathcal O}(p_0^4) \right]\nonumber\\
\iff 2 \kappa&=& 2 \; \nu\; p_0  + 2 \; (\nu-\sigma^2)\; p_0^2  
+ 4( \frac{2}{3}\; 
\nu\;  - \; \sigma^2 \; + \tau) \; p_0^3 + 
{\mathcal O}(p_0^4)
\label{reacrate2}
\end{eqnarray}
in terms of the frequency (\ref{freq}) of collisions on the catalysts $\nu$ and 
of the quantities

\begin{eqnarray}
\sigma^2&=&\lim_{t \to \infty}\;\frac{1}{t}\; \left(\langle N_t^2 \rangle-
\langle N_t \rangle^2\right) \label{sigma2} \\
\tau&=&\lim_{t \to \infty}\;\frac{1}{t}\;\left(\frac{1}{3}\;\langle N_t^3 
\rangle+\frac{2}{3}\;\langle N_t \rangle^3-\langle N_t^2 \rangle\langle N_t 
\rangle\right)\; .
\end{eqnarray}
The two further terms are given by
\begin{eqnarray}
\frac{\partial Q}{\partial \alpha}(1,1)
&=&-\lim_{t \to \infty}\frac{1}{t}\left[ 2\; p_0\; 
\langle N_t \rangle+2\;p_0^2\;\langle N_t \rangle+\frac{8}{3}\;p_0^3
\;\langle N_t \rangle \right.\nonumber \\
& &-4\;p_0^2\; \left(\langle N_t^2 \rangle-\langle N_t \rangle^2 \right)-8
\;p_0^3\;
\left(\langle N_t^2 \rangle-\langle N_t \rangle^2 \right) \nonumber \\
& & \left. +12\;p_0^3\; \left(\frac{1}{3}\;\langle N_t^3 \rangle
+\frac{2}{3}\;\langle N_t \rangle^3-\langle N_t^2 \rangle \;\langle N_t 
\rangle \right)+{\mathcal O}(p_0^4)\right]\nonumber \\
\iff \frac{\partial Q}{\partial \alpha}(1,1)&=&-2\kappa+2\;\sigma^2\;p_0^2
+4\;(\sigma^2\;-2\;\tau)\;p_0^3
+{\mathcal O}(p_0^4) \label{Qa}
\end{eqnarray}
and
\begin{eqnarray}
\frac{\partial Q}{\partial \beta}(1,1)=- \lambda &+&2\;p_0\;
\lim_{t \to \infty}\;\frac{1}{t}\;C_{\kappa \lambda}(t) \nonumber \\
&+&2\;p_0^2\;\lim_{t \to \infty}\;\frac{1}{t}
\left[\left(1+2\langle N_t \rangle\right)\;C_{\kappa \lambda}(t)
-D_{\kappa \lambda}(t)\right] \nonumber \\
&+&4\;p_0^3\;\lim_{t \to \infty}\;\frac{1}{t}\left[\left(\frac{2}{3}
+2\langle N_t \rangle+2\;\langle N_t \rangle^2
-\langle N_t^2 \rangle\right)\;C_{\kappa \lambda}(t)\right.\nonumber \\
& &\left. -(1+\langle N_t \rangle)\;D_{\kappa \lambda}(t)
+\frac{1}{3}E_{\kappa \lambda}(t)\right] 
+{\mathcal O}(p_0^4) \nonumber \\
\iff \frac{\partial Q}{\partial \beta}(1,1)&=&-\lambda
+\alpha_{\kappa \lambda}p_0+\beta_{\kappa \lambda}p_0^2+
\gamma_{\kappa \lambda}p_0^3+{\mathcal O}(p_0^4)
 \label{Qb}
\end{eqnarray}
where
\begin{eqnarray}
\lambda &=& \lim_{t\to\infty} \;\frac{1}{t}\; \langle \ln|\Lambda_t|
\rangle \label{lambda}\\
C_{\kappa \lambda}(t)&=&\langle N_t\;\ln|\Lambda_t| \rangle-\langle N_t 
\rangle
\langle \ln |\Lambda_t| \rangle \label{Ckl}\\
D_{\kappa \lambda}(t)&=&\langle N_t^2\;\ln|\Lambda_t| \rangle-\langle N_t^2 
\rangle \langle \ln |\Lambda_t| \rangle \\
E_{\kappa \lambda}(t)&=&\langle N_t^3\;\ln|\Lambda_t| \rangle-\langle N_t^3 
\rangle \langle \ln |\Lambda_t| \rangle \\
\alpha_{\kappa \lambda}&=&2\;\lim_{t \to \infty}\;\frac{1}{t}\;
C_{\kappa \lambda}(t) \label{alpha}\\
\beta_{\kappa \lambda}&=&2\;\lim_{t \to \infty}\;\frac{1}{t}
\left[\left(1+2\langle N_t \rangle\right)C_{\kappa \lambda}(t)
-D_{\kappa \lambda}(t)\right]
\label{beta}\\
\gamma_{\kappa \lambda}&=&4\;\lim_{t \to \infty}\;\frac{1}{t}
\left[\left(\frac{2}{3}+2\langle N_t \rangle+2\;\langle N_t\rangle^2 
-\langle N_t^2 \rangle\right)\;C_{\kappa \lambda}(t)\right. \nonumber \\
& &\left. -(1+\langle N_t \rangle)D_{\kappa \lambda}(t)
+\frac{1}{3}E_{\kappa \lambda}(t)\right]\; \label{gkl}.
\end{eqnarray}
For these coefficients, the index $\kappa$ refers to the reaction, 
related to the number of collisions on a catalyst $N_t$, while the 
index $\lambda$ refers to the dynamical instability, described by the 
stretching factor $\Lambda_t$ and the Lyapunov exponent $\lambda$.

  The reactive diffusion coefficient (\ref{reacdiffcoeff}) can also 
be expanded in powers of the reaction probability $p_0$ as
\begin{equation}
{\cal D}^{(r)} = {\cal D} + \Delta \; p_0 + {\mathcal O}(p_0^2)
\label{Drp0}
\end{equation}
in terms of the standard diffusion coefficient
\begin{equation}
{\cal D} = \lim_{t\to\infty} \frac{1}{2\; d\; t}\; \langle ({\bf r}_t-{\bf
r}_0)^2\rangle
\label{diffcoeff}
\end{equation}
and of the coefficient
\begin{equation}
\Delta = - \; \lim_{t\to\infty} \frac{1}{d\; t}\; \left[ \langle N_t({\bf r}_
t-{\bf r}_0)^2\rangle - \langle N_t\rangle \langle({\bf r}_t-{\bf
r}_0)^2\rangle \right]
\label{Delta}
\end{equation}
which characterizes the correlation between the transport by diffusion and 
the reaction.

We notice that, in the limit $p_0\to 0$, the reaction rate (\ref{reacrate2}) 
vanishes while the reactive diffusion coefficient (\ref{reacdiffcoeff}) tends 
to the diffusion coefficient (\ref{diffcoeff}), as expected.

Using (\ref{Qa}) and (\ref{Qb}), the Hausdorff dimension (\ref{hausdim}) 
becomes
\begin{eqnarray}
D_{\rm H}({\bf k})&=&1+\frac{{\cal D}^{\rm (r)}k^2}{\lambda\;\left[1-
\frac{\alpha_{\kappa \lambda}}{\lambda}\;p_0-
\frac{\beta_{\kappa \lambda}}{\lambda}\;p_0^2-
\frac{\gamma_{\kappa \lambda}}{\lambda}\;p_0^3-
2\;\frac{\sigma^2}{\lambda}\;p_0^2-4\;\frac{\sigma^2}{\lambda}\;p_0^3
+8\;\frac{{\tau}}{\lambda}\;p_0^3+{\mathcal O}(p_0^4)\right]}
+{\mathcal O}(k^4)
 \nonumber \\
&=&1+\frac{{\cal D}^{\rm (r)}k^2}{\lambda}\;\left[1+
\frac{\alpha_{\kappa \lambda}}{\lambda}\;p_0+
2\;\frac{\sigma^2}{\lambda}\;p_0^2+
\frac{\beta_{\kappa \lambda}}{\lambda}\;p_0^2+
\frac{\alpha_{\kappa \lambda}^2}{\lambda^2}\;p_0^2+
4\;\frac{\sigma^2}{\lambda}\;p_0^3
-8\;\frac{{\tau}}{\lambda}\;p_0^3 \right.\nonumber \\ 
& & \left. +\frac{\gamma_{\kappa \lambda}}{\lambda}\;p_0^3+
2\;\frac{\alpha_{\kappa \lambda}\beta_{\kappa \lambda}}{\lambda^2}\;p_0^3
+4\;\frac{\alpha_{\kappa \lambda}\sigma^2}{\lambda^2}\; p_0^3
+\frac{\alpha_{\kappa \lambda}^3}{\lambda^3}\; p_0^3
+{\mathcal O}(p_0^4)\right]+{\mathcal O}(k^4) \; .
\label{hausdim2}
\end{eqnarray}
or
\begin{equation}
D_{\rm H}({\bf k})=1+\left[\frac{{\cal D}^{\rm (r)}}{\lambda}
+{\mathcal O}(p_0)\right]\;k^2
+{\mathcal O}(k^4)\; .
\end{equation}
We immediately see that for $p_0 \to 0$, Eq.\ (\ref{hausdim2}) reduces to 
the expression (\ref{diff}) obtained for the diffusive case.

We are now going to test this formula for two examples: 
a reactive triadic multibaker and a reactive two-dimensional periodic 
Lorentz gas.
%%%%%%%%%%%%%%%%%%%%%%%%%%%%%%%%%%%%%%%%%%%%%%%%%%%%%%%%%%%%%%%%%%%%%%%%%%%%%%%
\section{The reactive triadic multibaker map}
\label{triadic}
\subsection{Description of the model}
\begin{figure}[ht]
\centerline{\epsfxsize=10 truecm \epsfbox{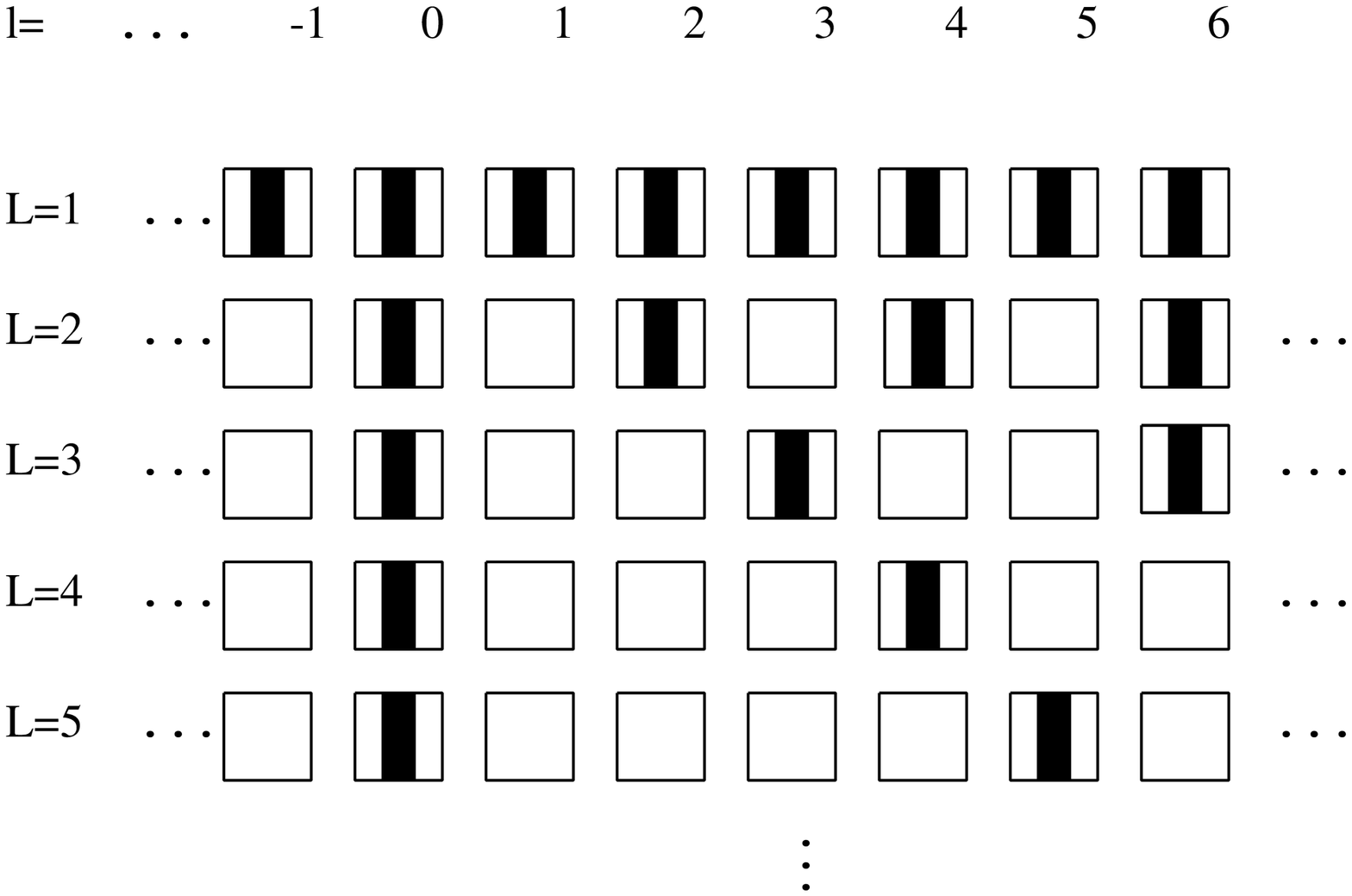}}
\caption{Geometry of the reactive triadic multibaker map for various values 
of the distance $L$ between the catalysts.}\label{multi}
\end{figure}

The multibaker map is a simple model used to mimic a diffusive motion 
\cite{Gilb1,Gasp2}. In the case of the triadic symmetric multibaker, 
we consider an infinite one-dimensional chain of squares,
\begin{equation}
{\mathcal S}_l=\{(x,y,l):\;0\le x\le 1,\;0\le y\le 1\} \; ,
\end{equation}
where $l \in \hbox{\mb Z}$ is the label of the square. A point particle 
performs jumps from square to square, according to the transition rule 
$l \to (l-1)$ if $x \le 1/3$, $l \to l$ if $1/3<x \le 2/3$ and $l \to (l+1)$ 
if $x>2/3$. In order to introduce a reaction, the point particle carries 
a color $A$ or $B$, and the central part $]\frac{1}{3},\frac{2}{3}]$ of 
one square over $L$ is catalytic: if the point particle is in this portion of 
the square, it changes its color with a probability $p_0$ 
(see Fig.\ \ref{multi}). The map of the model is thus given by 
\begin{equation}
\bphi(x,y,l,c)=\cases{
(3x,\frac{y}{3},l-1,c)\; , & $0 \le x \le 1/3$ \cr
(3x-1,\frac{y+1}{3},l,\varepsilon(l)\;c)\; , & $1/3 < x \le 2/3$\cr
(3x-2,\frac{y+2}{3},l+1,c)\; , & $2/3 < x \le 1$\cr}
\label{bak}
\end{equation}
where $c$ is a discrete variable taking the values $+1$ or $-1$ when the 
color of the point particle is respectively $A$ or $B$, and
\begin{equation}
\varepsilon(l)=\cases{
-1 & if $l$ is a catalytic square, with a probability $p_0$ ,\cr
+1 & otherwise . \cr}
\end{equation}
We consider the probability density $q(x,y,l,c)$ which evolves under the  
Frobenius-Perron equation
\begin{equation}
q_{t+1}(x,y,l,c)=\cases{
q_t\left(\frac{x}{3},3y,l+1,c\right)\; , & $0 \le y \le 1/3$\cr
[1-p_0(l)]\; q_t\left(\frac{x+1}{3},3y-1,l,c\right)+p_0(l)\; 
q_t\left(\frac{x+1}{3},3y-1,l,-c\right)\; , & $1/3 < y \le 2/3$\cr
q_t\left(\frac{x+2}{3},3y-2,l-1,c\right)\; , & $2/3 < y \le 1$\cr}
\end{equation}
where
\begin{equation}
p_0(l)=\cases{
p_0 & if $l$ is a catalytic square , \cr
0   & otherwise .\cr}
\end{equation}
The Kolmogorov-Sinai entropy per iteration (\ref{KS}) of the reactive triadic 
multibaker is given by
\begin{equation}
h_{\rm KS}= \ln 3 + \frac{p_0}{3L}\;
\ln\frac{1}{p_0} + \frac{1-p_0}{3L}\; \ln\frac{1}{1-p_0} 
\label{KSbak}
\end{equation}
because the frequency (\ref{freq}) is equal to the ratio of
the area of a reactive rectangle to the area of $L$ squares: $\nu=1/3L$. We 
notice that the result (\ref{KSbak}) can also be obtained by using a 
generating partition which establishes the equivalence between the reactive 
multibaker map (\ref{bak}) and a Markov chain (see Ref.\ \cite{Gaspbook}).

If we define the difference between the densities of the colors $A$ and $B$,
\begin{equation} 
g_t(x,y,l)=q_t(x,y,l,+1)-q_t(x,y,l,-1)\; ,
\end{equation}
 we obtain the following equation for its evolution
\begin{equation}
g_{t+1}(x,y,l)=\cases{
g_t\left(\frac{x}{3},3y,l+1\right)\; , & $0 \le y \le 1/3$ \cr
[1-2p_0(l)]\; g_t\left(\frac{x+1}{3},3y-1,l\right)\; , & $1/3 <  y 
\le 2/3$ \cr
g_t\left(\frac{x+2}{3},3y-2,l-1\right)\; , & $2/3 < y \le 1 $\cr}
\label{bakop}
\end{equation}
which is the form of the reactive operator (\ref{opreac}) in the case of the 
present multibaker model.
%%%%%%%%%%%%%%%%%%%%%%%%%%%%%%%%%%%%%%%%%%%%%%%%%%%%%%%%%%%%%%%%%%%%%%%%%%%%

\subsection{Relaxation rates and eigenmodes of the reactive operator}

We restrict our description to a unit cell of length $L$, $l=0$ being 
the catalytic square, and we choose quasiperiodic boundary conditions,
assuming that the solution of the equation (\ref{bakop}) is quasiperiodic 
on the chain with a wavenumber $k$. Moreover, we suppose exponentially 
decaying solutions, with decaying factor $\chi(k)=\exp s_k$ where 
$|\chi(k)|\le 1$ or ${\rm Re}s_k \le 0$
\begin{equation}
g_t(x,y,l)\sim\chi^t\;\exp(ikl).
\end{equation}
The following operator $\hat{Q}_k$ governs the time 
evolution of $g_t(x,y,l)$ \cite{Gasp2}:
\begin{equation}
\hat{Q}_k \equiv \cases{
g_{t+1}(x,y,0) & $=\theta (\frac{1}{3}-y)g_t(\frac{x}{3},3y,1)
+(1-2p_0)\theta (y-\frac{1}{3})\theta (\frac{2}{3}-y)g_t(\frac{x+1}
{3},3y-1,0)
$\cr
 & $+e^{-ikL}\theta (y-\frac{2}{3})g_t(\frac{x+2}{3},3y-2,L-1)$\cr
g_{t+1}(x,y,1)& $=\theta (\frac{1}{3}-y)g_t(\frac{x}{3},3y,2)
+\theta (y-\frac{1}{3})\theta (\frac{2}{3}-y)g_t(\frac{x+1}{3},3y-1,1)
$\cr
 & $+\theta (y-\frac{2}{3})g_t(\frac{x+2}{3},3y-2,0)$\cr
 & $\vdots$\cr
g_{t+1}(x,y,L-2)& $=\theta (\frac{1}{3}-y)g_t(\frac{x}{3},3y,L-1)
+\theta (y-\frac{1}{3})\theta (\frac{2}{3}-y)g_t(\frac{x+1}{3},3y-1,L-2)
$\cr
 & $+\theta (y-\frac{2}{3})g_t(\frac{x+2}{3},3y-2,L-3)$\cr
g_{t+1}(x,y,L-1)& $=-e^{ikL}\theta (\frac{1}{3}-y)g_t(\frac{x}{3},3y,0)
+\theta (y-\frac{1}{3})\theta (\frac{2}{3}-y)g_t(\frac{x+1}{3},3y-1,L-1)
$\cr
 & $+\theta (y-\frac{2}{3})g_t(\frac{x+2}{3},3y-2,L-2)$.\cr}
\label{operator}
\end{equation}
We now want to obtain the eigenvalues and the eigenstates of this operator
\begin{equation}
\hat{Q}_k\{ \Psi(x,y,l)\}_{l=0}^{l=L-1}=\chi\{ \Psi(x,y,l)\}_{l=0}^
{l=L-1}\; .
\end{equation}
We suppose that the leading eigenstates are uniform along the unstable 
direction $x$
\begin{equation}
\Psi(x,y,l)={\mathcal Y}(y,l)\; .
\end{equation}
If we now cumulate over $x$ and $y$, we get
\begin{equation}
G_t(x,y,l)=\int_0^x dx'\int_0^ydy'\Psi(x',y',l)\equiv x\;  C(y,l)
\end{equation}
where
\begin{equation}
C(y,l)=\int_0^y dy' {\mathcal Y}(y',l)
\end{equation}
is the reactive cumulative function (\ref{cumulreac}) for the multibaker.
We replace $g_t(x,y,l)$ by ${\mathcal Y}(y,l)$ and $g_{t+1}(x,y,l)$ by 
$\chi{\mathcal Y}(y,l)$ in Eq.\ (\ref{operator}) and we integrate over 
the interval $[0,y]$. We obtain
\begin{eqnarray}
C(y,0) &=&\cases{
\frac{1}{3\chi}\;C(3y,1)\; , & $0\le y \le 1/3$\cr
\frac{1}{3\chi}\;[C(1,1)+(1-2p_0)C(3y-1,0)]\; ,& $1/3< y \le 2/3$\cr
\frac{1}{3\chi}\;[C(1,1)+(1-2p_0)C(1,0)+e^{-ikL}C(3y-2,L-1)] \; , & $2/3< y 
\le 1$\cr}
\nonumber\\
C(y,1) &=&\cases{
\frac{1}{3\chi}\;C(3y,2) \; , & $0\le y \le 1/3$\cr
\frac{1}{3\chi}\;[C(1,2)+C(3y-1,1)] \; , &$ 1/3< y \le 2/3$\cr
\frac{1}{3\chi}\;[C(1,2)+C(1,1)+C(3y-2,0)] \; , & $2/3< y \le 1$\cr}
\nonumber\\
&\vdots & \nonumber \\
C(y,L-2) &=&\cases{
\frac{1}{3\chi}\;C(3y,L-1) \; , & $0\le y \le 1/3$\cr
\frac{1}{3\chi}\;[C(1,L-1)+C(3y-1,L-2)] \; , &$ 1/3< y \le 2/3$\cr
\frac{1}{3\chi}\;[C(1,L-1)+C(1,L-2)+C(3y-2,L-3)] \; , & $2/3< y \le 1$\cr}
\nonumber\\
C(y,L-1) &=&\cases{
\frac{e^{ikL}}{3\chi}\;C(3y,0) \; , & $0\le y \le 1/3$\cr
\frac{1}{3\chi}\;[e^{ikL}\;C(1,0)+C(3y-1,L-1)] \; , & $1/3< y \le 2/3$\cr
\frac{1}{3\chi}\;[e^{ikL}\;C(1,0)+C(1,L-1)+C(3y-2,L-2)] \; , & $2/3< y 
\le 1$\cr}.
\label{Cequ}
\end{eqnarray}

The eigenvalues are obtained by setting $y=1$ in Eq.\ (\ref{Cequ}), which 
leads to the eigenvalue equation
\begin{equation}
\pmatrix{
(1-2p_0-3\chi) & 1 & 0 & \cdots & 0 & e^{-ikL}\cr
1 & (1-3\chi) & 1 & \cdots & 0 & 0\cr
0 & 1 & (1-3\chi) & \cdots & 0 & 0\cr
\vdots & \vdots & \vdots & \ddots & \vdots & \vdots\cr
0 & 0 & 0 & \cdots & (1-3\chi) & 1\cr
e^{ikL} & 0 & 0 & \cdots & 1 & (1-3\chi)\cr} 
\pmatrix{
C(1,0)\cr
C(1,1)\cr
C(1,2)\cr
\vdots\cr
C(1,L-2)\cr
C(1,L-1)\cr}=0\; . \label{eigenequ}
\end{equation}
We thus have to solve the determinant of an Hermitian matrix to find $\chi(k)$.
For this purpose, we assume a solution of the form
\begin{equation}
C(1,l) = A\; e^{+i\theta l} + B\; e^{-i\theta l} \; .
\end{equation}
After substitution in the system (\ref{eigenequ}), we find that
\begin{equation}
\chi= \frac{1+2\cos\theta}{3} 
\label{chi}
\end{equation}
and 
\begin{equation}
\cos kL = \cos\theta L + p_0 \; \frac{\sin\theta L}{\sin \theta}.
\label{determinant}
\end{equation}
For the present reactive system, the relaxation rate $s(k)=\ln\chi(k)$ depends 
on the wavenumber $k$ as $s(k)=-2\kappa-{\cal D}^{\rm (r)}k^2
+{\mathcal O}(k^4)$, where $\kappa$ is the reaction rate and 
${\cal D}^{\rm (r)}$ is the reactive diffusion coefficient.

Solving Eq.\ (\ref{determinant}) for its leading root, we can thus obtain the 
reaction rate and the reactive diffusion coefficient either in powers of the 
transition probability $p_0$:
\begin{eqnarray}
\kappa &=& \frac{p_0}{3L} + \left( \frac{1}{L^2}-\frac{1}{3}\right) 
\frac{p_0^2}{6} + {\mathcal O}(p_0^3)\; , \label{ratep0}\\
{\cal D}^{\rm (r)} &=& \frac{1}{3} + \frac{2p_0}{9L} + {\mathcal O}(p_0^2)\; ,
\label{diffreacp0}
\end{eqnarray}
or in the asymptotic limit $L\to\infty$:
\begin{eqnarray}
\kappa &=& \frac{\pi^2}{6L^2} - \frac{2\pi^2}{3p_0L^3} +
{\mathcal O}\left(\frac{1}{L^4}\right)\; , \label{rateL}\\
{\cal D}^{\rm (r)} &=& \frac{\pi^2}{3p_0L} +
{\mathcal O}\left(\frac{1}{L^2}\right)\; . \label{diffreacL}
\end{eqnarray}
In the limit $p_0\to 0$, Eq.\ (\ref{ratep0}) shows that the reaction rate 
vanishes linearly with the transition probability and Eq.\ (\ref{diffreacp0}) 
that the reactive diffusion coefficient tends to the purely diffusive one 
${\cal D}=1/3$, as expected.  On the other hand, Eq.\ (\ref{rateL}) shows that 
the reaction rate decreases as $L^{-2}$ when the
catalytic sides are diluted in the limit $L\to\infty$, as expected for a
diffusion-controled reaction in a one-dimensional system \cite{Gasp2}.  
Moreover, the vanishing of the reactive diffusion coefficient 
(\ref{diffreacL}) as $L^{-1}$ for
$L\to\infty$ is consistent with the results previously obtained in Ref.\  
\cite{Gasp2}.  We notice the remarkable property that, in the 
diffusion-controled regime of the reaction, the
leading term of the asymptotic expression for the reaction rate (\ref{rateL}) 
does not depend on the transition probability $p_0$, which only appears in 
the next terms. Therefore, the leading term of (\ref{rateL}) no longer 
expresses the property that the reaction rate vanishes in the limit 
$p_0\to 0$, as shown by Eq.\ (\ref{ratep0}).  This is
consistent with the fact that the reaction rate is largely independent of 
the local reaction rate at the catalysts in the diffusion-controled regime, 
as already noticed in Refs.\ \cite{Claus2,Claus1}.

Let us consider more explicitly the simple case $L=1$. In this case, 
Eq.\ (\ref{eigenequ}) reduces to 
\begin{equation}
3\chi-2\cos k-1+2p_0=0
\end{equation}
which gives the solution
\begin{equation}
\chi(k)=\frac{1-2p_0+2\cos k}{3} \label{chiL1} \; .
\end{equation}
For $k\ll 1$, $\chi(k)$ is then given by 
\begin{eqnarray}
\chi(k)&\simeq&e^{-2\kappa}e^{-{\cal D}^{\rm (r)}k^2} \nonumber \\
       &\simeq&\chi_0\;(1-{\cal D}^{\rm (r)}k^2)\; .
\end{eqnarray}
Expanding (\ref{chiL1}) in powers of $k$ up to the second degree, we obtain 
\begin{equation}
\chi\simeq\frac{3-2p_0}{3}\;\left(1-\frac{1}{3-2p_0}k^2\right)
\label{chi0}
\end{equation}
which allows us to identify $\chi_0=\frac{3-2p_0}{3}=e^{-2\kappa}$ and 
${\cal D}^{\rm (r)}=\frac{1}{3-2p_0}$. 

The dependence of the reactive 
diffusion coefficient ${\cal D}^{\rm (r)}$ on $p_0$ is shown in Fig.\ 
\ref{DrvsL} for different values of the chain length $L$. The value of 
the diffusion coefficient $\cal D$ is indicated for comparison. 
%%%%%%%%%%%%%%%%%%%%%%%%%%%%%%%%%%%%%%%%%%%%%%%%%%%%%%%%%%%%%%%%%%%%%%%%%%%%%

\subsection{Fractal dimension of the cumulative functions of the eigenmodes}

\begin{figure}[ht]
\centerline{\epsfxsize=10 truecm \epsfbox{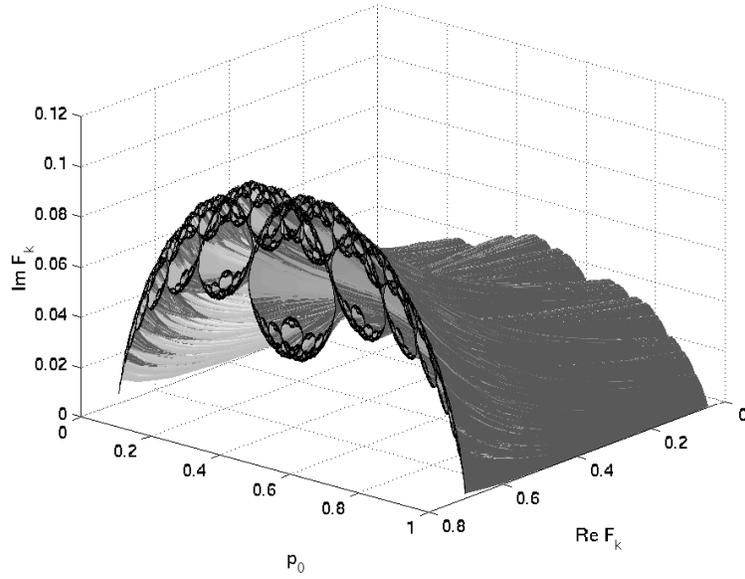}}
\caption{Three-dimensional representation of the cumulative functions of the 
reactive eigenmodes of the triadic multibaker; $({\rm Re}F_k,{\rm Im}F_k)$ are 
represented in the complex plane as a function of the probability of reaction 
$p_0$, in the case $L=1$, $k=0.1$.}\label{modevsp}
\end{figure}

\begin{figure}[ht]
\centerline{\epsfxsize=10 truecm \epsfbox{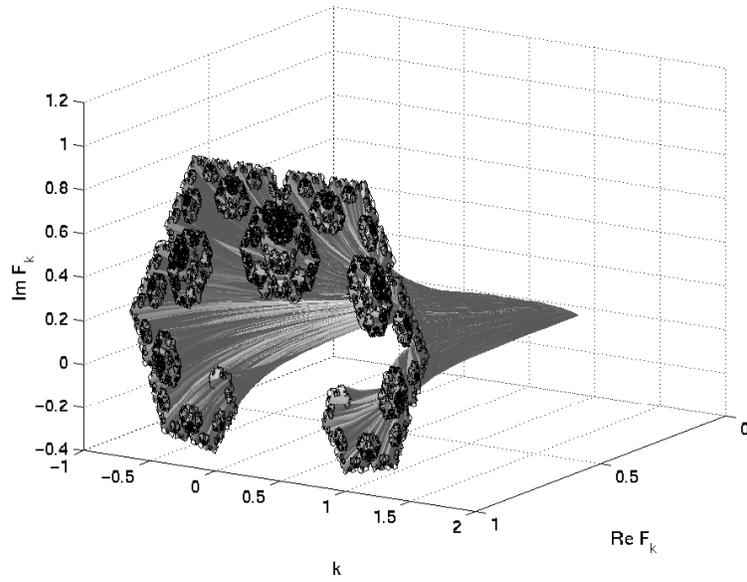}}
\caption{Three-dimensional representation of the cumulative functions of the 
reactive eigenmodes of the triadic multibaker; $({\rm Re}F_k,{\rm Im}F_k)$ are 
represented in the complex plane as a function of the wavenumber 
$k$, in the case $L=1$, $p_0=0.3$.}\label{modevsk}
\end{figure}

We now want to study the cumulative functions $C(y,l)$, solutions of Eq.\
(\ref{Cequ}), and more specifically to calculate their dimension. 
These functions are represented in the complex plane in the case $L=1$ in 
Figs.\ \ref{modevsp} and \ref{modevsk}. Fig.\ \ref{modevsp} shows their 
dependence on the reaction probability $p_0$, the wavenumber being fixed, 
$k=0.1$. Their dependence on the wavenumber for $p_0=0.3$ is shown in Fig.\ 
\ref{modevsk}.
As explained in Section \ref{theory}, these complex cumulative functions can 
be approximated by a polygon at a given time $t$, which means at a certain 
iteration step $n$ in this 
discrete case. This concept will here be applied 
using a symbolic dynamics.  

\begin{figure}[ht]
\centerline{\epsfxsize=10 truecm \epsfbox{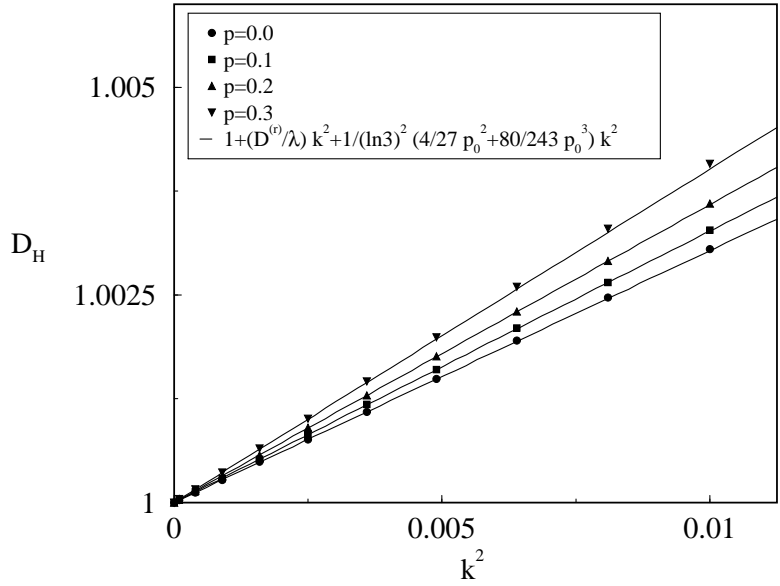}}
\caption{Reactive triadic multibaker: dependence of the Hausdorff dimension 
$D_{\rm H}$ of the cumulative functions of the reactive eigenmodes on 
the square of the wavenumber $k^2$, for $L=1$ and  for different values of 
the reaction probability $p_0$. The 
symbols are obtained by numerical resolution of  Eq.\ (\ref{bakdh}) and the 
solid line is the analytic solution (\ref{bakdim}) obtained in the limit 
$p_0 \ll 1$.}
\label{Dhvsk2}
\end{figure}

The symbolic dynamics is constructed by partitioning the phase space
of the multibaker maps depicted in Fig.\ \ref{multi}.  Each of the 
$L$ squares of the chain is divided in three 
parts for $x\in[0,\frac{1}{3}]$, $x\in]\frac{1}{3},\frac{2}{3}]$ or 
$x\in]\frac{2}{3},1]$. 
These $3L$ rectangles are successively assigned the symbols $\omega_n
\in \{0,1,\cdots,3L-2,3L-1\}$ from left to right along 
the chain. At the $n^{\rm th}$ step of the construction described in 
Subsection \ref{reacmodes}, the cumulative function is a polygon in the complex 
plane with its sides corresponding to a subrectangle associated with a sequence 
of $n$ symbols $\{\omega_{n-1}\omega_{n-2}\cdots\omega_1\omega_0\}$. Moreover, 
we notice that the sequence of symbols also determines the square $l$ where the 
point is located along the chain. At the next steps of the construction, the 
cumulative function keeps its values already obtained at the borders of each of 
the aforementioned subrectangles.  We can thus denote the values of the 
cumulative function at the lower and upper borders $y_{\pm}(\omega_{n-1},
\omega_{n-2},\cdots,\omega_1,\omega_0)$ of one of the subrectangles in the
square $l$ more shortly by
\begin{equation}
C_{\pm}(\omega_{n-1},\omega_{n-2},\dots,\omega_1,\omega_0) =
C\left[y_{\pm}(\omega_{n-1},\omega_{n-2},\dots,\omega_1,\omega_0),l\right].
\end{equation}
The complex vector defining the side of the polygon at the $n^{\rm th}$ step 
of the construction is thus given by
\begin{equation}
\Delta C(\omega_{n-1},\omega_{n-2},\dots,\omega_1,\omega_0)=
C_+(\omega_{n-1},\omega_{n-2},\dots,\omega_1,\omega_0)-
C_-(\omega_{n-1},\omega_{n-2},\cdots,\omega_1,\omega_0).
\end{equation}
The succession of two symbols $\omega_1\omega_0$ corresponds to a  
transition in the multibaker phase-space, with which a certain 
probability is associated . Using (\ref{Cequ}), we can construct 
a $3L\times 3L$ transition 
matrix $\mathsf M$ of elements $M_{\omega_1 \omega_0}$ describing the 
factor connecting the portion of square corresponding to $\omega_0$ to 
the one corresponding to $\omega_1$. Using this matrix, we can describe 
the time evolution of the intervals 
$\Delta C(\omega_{n-1},\omega_{n-2},\dots,\omega_1,\omega_0)$ as 
\begin{equation}
\Delta C(\omega_{n-1},\omega_{n-2},\dots,\omega_1,\omega_0)=
M_{\omega_{n-1} \omega_{n-2}}M_{\omega_{n-2} \omega_{n-3}} \cdots
M_{\omega_2 \omega_1}M_{\omega_1 \omega_0}
\Delta C(\omega_0)\; .
\end{equation}
We define
\begin{eqnarray}
\Gamma_n^D(C)&=&\sum_{\omega_0 \cdots \omega_{n-1}}\left| \Delta 
C(\omega_{n-1},\omega_{n-2},\dots,\omega_1,\omega_0) \right|^D \nonumber \\
&=&\sum_{\omega_0 \cdots \omega_{n-1}}
\left|M_{\omega_{n-1}\omega_{n-2}} \right|^D \cdots
\left|M_{\omega_{1}\omega_0} \right|^D
\left|\Delta C(\omega_0) \right|^D\\
&=&\sum_{\omega_0\omega_{n-1}}
({\mathcal M}^{n-1})_{\omega_{n-1}\omega_0}\left|\Delta C(\omega_0) \right|^D
\label{Gamma}
\end{eqnarray}
where the matrix $\mathcal M$ is composed of the elements:
\begin{equation}
{\mathcal M}_{\omega \omega'}=\left|M_{\omega \omega'} \right|^D \; .
\end{equation}
Supposing that this matrix admit the spectral decomposition, 
${\mathcal M}=\sum_{\mu}|\mu\rangle \; \mu\; \langle\mu|$, Eq.\ (\ref{Gamma}) 
can be rewritten as 
\begin{equation}
\Gamma_n^D(C)=\sum_{\mu} \; \langle {\bf u} |\mu\rangle \; \mu^{n-1} \; 
\langle \mu| {\bf v}\rangle \label{reGamma}
\end{equation}
where $\bf u$ denotes the vector of unit elements $u_{\omega}=1$ and $\bf v$ 
the vector with the elements $v_{\omega}=|\Delta C(\omega)|^D$.  According to the 
definition of the Hausdorff dimension, for $n \to \infty$, the quantity 
(\ref{reGamma}) must converge to a finite non-vanishing value if $D=D_{\rm H}$. 
This will only be the case if $\mu=1$. $D_{\rm H}$ is thus solution of the 
equation obtained by imposing $1$ as eigenvalue of the matrix ${\mathcal M}$. 
Let us consider explicitly the case $L=1$. We have to calculate the $3\times 3$ 
determinant
\begin{equation}
\left|
\matrix{
\left|\frac{1}{3 \chi} \right|^{D_{\rm H}}-1 & \left|\frac{1}{3 \chi} 
\right|^{D_{\rm H}} 
& \left|\frac{1}{3 \chi} \right|^{D_{\rm H}} \cr
\left|\frac{1-2p_0}{3 \chi} \right|^{D_{\rm H}} & \left|\frac{1-2p_0}{3 \chi} 
\right|^{D_{\rm H}}-1 & \left|\frac{1-2p_0}{3 \chi} \right|^{D_{\rm H}} \cr
\left|\frac{1}{3 \chi} \right|^{D_{\rm H}} & \left|\frac{1}{3 \chi} 
\right|^{D_{\rm H}} & 
\left|\frac{1}{3 \chi} \right|^{D_{\rm H}}-1 \cr}
\right|=0.
\end{equation}
This gives the equation 
\begin{equation}
\left(\frac{1-2p_0}{3 \chi} \right)^{D_{\rm H}}+2\left(\frac{1}{3 \chi} 
\right)^{D_{\rm H}}-1=0 \; .
\label{bakdh}
\end{equation}
We will solve (\ref{bakdh}) in the limit case $p_0 \to 0$, using the 
expression of $\chi$ in the case $L=1$ (\ref{chiL1}).
For $p_0=0$, we have $D_{\rm H}^{(0)}=\frac{\ln 3}{\ln(1+2\cos k)}$. 
On the other hand, the Hausdorff dimension takes the unit value for $k=0$.
Substituting the expansion $D_{\rm H}=1+ak^2+{\cal O}(k^4)$ in 
Eq.\ (\ref{bakdh}) and expanding in powers of the wavenumber $k$, we can solve 
for the unknown coefficient $a$ and obtain
\begin{equation}
D_{\rm H} = 1 + \frac{k^2}{(3-2p_0)\ln(3-2p_0)-(1-2p_0)\ln(1-2p_0)}
+{\cal O}(k^4)\; .
\end{equation}
Now expanding in powers of the transition probability $p_0$, we finally obtain
\begin{equation}
D_{\rm H}=1+\frac{1}{\ln 3}\;\frac{1}{3}\left(1+\frac{2}{3}p_0
+\frac{4}{9}p_0^2+
\frac{8}{27}p_0^3\right)k^2+\frac{1}{(\ln 3)^2}\left(\frac{4}{27}p_0^2+
\frac{80}{243}p_0^3\right)\; k^2 +{\mathcal O}(p_0^4k^2) +{\mathcal O}(k^4)
\label{bakdim}
\end{equation} 
We can again compare this expression with the general result (\ref{hausdim2}) 
of Section \ref{theory}. 
For the triadic multibaker, the Lyapunov exponent is given by $\lambda=\ln 3$.
Moreover, the stretching factor
$\ln |\Lambda_t| $ is constant in the triadic multibaker, which implies that 
the  coefficients $C_{\kappa \lambda}$, $D_{\kappa \lambda}$,
$E_{\kappa \lambda}$ and therefore $\alpha_{\kappa \lambda}$, 
$\beta_{\kappa \lambda}$, $\gamma_{\kappa \lambda}$ defined 
in Eqs.\ (\ref{Ckl})-(\ref{gkl}) are vanishing.
Whereupon, Eq.\ (\ref{hausdim2}) is here reduced to 
\begin{equation}
D_{\rm H}(k)=1+\frac{{\cal D}^{\rm (r)}k^2}{\lambda}\;\left[1+2\;
\frac{\sigma^2}
{\lambda}\;p_0^2+4\;\frac{\sigma^2}{\lambda}\;p_0^3
-8\;\frac{{\tau}}{\lambda}\;p_0^3+{\mathcal O}(p_0^4)\right]
+{\mathcal O}(k^4)\; .
\label{hausdim3}
\end{equation}
Using the expansion of the reactive diffusion coefficient up to the third 
order in $p_0$
\begin{equation}
{\cal D}^{\rm (r)}=\frac{1}{3}\left[1+\frac{1}{3}p_0+\frac{4}{9}p_0^2
+\frac{8}{27}p_0^3+{\mathcal O}(p_0^4)\right]\; ,
\end{equation}
Eq.\ (\ref{hausdim3}) becomes 
\begin{eqnarray}
D_{\rm H}(k)&=&1+\frac{1}{\ln 3}\;\frac{1}{3}\left(1+\frac{1}{3}p_0
+\frac{4}{9}p_0^2
+\frac{8}{27}p_0^3\right)k^2+\frac{1}{(\ln 3)^2}\left(\frac{1}{3}
+\frac{2}{9}\;p_0\right)
\;2\sigma^2\;p_0^2\;k^2 \nonumber \\
& & +\frac{1}{(\ln 3)^2}\frac{1}{3}\;4\;\sigma^2\;p_0^3\;k^2
-\frac{1}{(\ln 3)^2}\;\frac{1}{3}\;8\;\tau\;p_0^3\;k^2 +{\mathcal
O}(p_0^4 k^2)+{\mathcal O}(k^4) \;.
\label{hausdim4}
\end{eqnarray}
The second term of the right members of (\ref{bakdim}) and (\ref{hausdim4}) 
are equal. They correspond to $\frac{{\cal D}^{\rm (r)}k^2}
{\lambda}$. Let us now compare the other corrections in $p_0$. The only 
correction in $p_0^2$ in (\ref{hausdim4}) is 
$\frac{1}{(\ln 3)^2}\;\frac{1}{3}\;2\;\sigma^2\;p_0^2\;k^2$. We can evaluate 
$\sigma^2=\lim_{t \to \infty}\;\frac{1}{t}\;(\langle N_t^2 \rangle-\langle 
N_t \rangle^2)$ in the case of the triadic reactive multibaker $L=1$. Indeed 
here, colliding on a catalyst corresponds to being in the region 
$x\in ]\frac{1}{3},\frac{2}{3}]$. The mean number of collisions on a 
catalyst per unit time is then 
\begin{equation}
\frac{1}{t}\;\langle N_t\rangle =\frac{1}{3} \cdot 0+ \frac{1}{3} \cdot 1+
\frac{1}{3} \cdot 0=\frac{1}{3}
\end{equation}
and 
\begin{equation}
\frac{1}{t}\;\langle N_t^2\rangle=\frac{1}{3} \cdot 0^2+ \frac{1}{3} 
\cdot 1^2+\frac{1}{3} \cdot 0^2=\frac{1}{3}
\end{equation}
which finally gives us 
\begin{equation}
\sigma^2=\frac{1}{3}-\left(\frac{1}{3}\right)^2=\frac{2}{9} \; .
\end{equation}
The coefficient of the correction in $p_0^2$ is then equal to 
$\frac{1}{(\ln 3)^2}\;\frac{4}{27}$, which is exactly the coefficient 
obtained in (\ref{bakdim}). 
The correction in $p_0^3$ in (\ref{hausdim4}) contains three different 
contributions: $\frac{1}{(\ln 3)^2}\frac{4}{9}\;\sigma^2\;p_0^3\;k^2
+\frac{1}{(\ln 3)^2}\frac{4}{3}\;\sigma^2\;p_0^3\;k^2
-\frac{1}{(\ln 3)^2}\;\frac{8}{3}\;\tau\;p_0^3\;k^2$. Again we can evaluate
$\tau$
\begin{eqnarray} 
\tau&=&\lim_{t \to \infty}\;\frac{1}{t}\;\left(\frac{1}{3}\;
\langle N_t^3 \rangle
+\frac{2}{3}\;\langle N_t \rangle^3-\langle N_t^2 \rangle\langle N_t 
\rangle\right)
\nonumber \\
&=&\frac{1}{3}\cdot\frac{1}{3}+\frac{2}{3}\cdot \frac{1}{27}
-\frac{1}{3}\cdot\frac{1}{3}=\frac{2}{81}\; .
\end{eqnarray}
We obtain $\frac{1}{(\ln 3)^2}\frac{80}{243}$, which is the coefficient 
expected for the correction in $p_0^3$. 
Accordingly, we have confirmed that Eq.\ (\ref{hausdim2}) for the Hausdorff 
dimension of the reactive modes is valid for the reactive multibaker map.
The validity of the approach is furthermore tested in Fig.\ \ref{Dhvsk2}, 
where the Hausdorff diemansion $D_{\rm H}$ is plotted versus the square of 
the wavenumber $k^2$. 
The symbols are obtained numerically by solving Eq.\ (\ref{bakdh}) and 
compared with the expression (\ref{bakdim}). The agreement is good up 
to $p_0=0.3$. We notice that, beyond this value, further terms would be 
required in the Taylor expansion (\ref{bakdim}).
%%%%%%%%%%%%%%%%%%%%%%%%%%%%%%%%%%%%%%%%%%%%%%%%%%%%%%%%%%%%%%%%%%%%%%%%%%%%%%
\section{The 2D reactive periodic Lorentz gas}
\label{lorentz}
\begin{figure}[ht]
\centerline{\epsfxsize=10 truecm \epsfbox{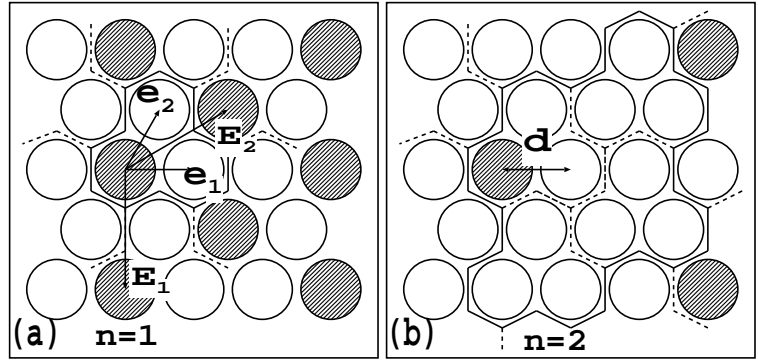}}
\caption{Two-dimensional reactive periodic Lorentz gas. 
Elementary cell of the superlattice of the catalysts (a) in the case $n=1$; 
(b) in the case $n=2$. }
\label{fundamental}
\end{figure}

We are now going to consider as second example a two-dimensional reactive 
periodic Lorentz gas. This model has been introduced in 
Ref.\ \cite{Claus1}. In this system, a point particle, carrying a color $A$ 
or $B$ undergoes elastic collisions on hard disks fixed in the plane and 
forming a regular triangular lattice characterized by the lattice fundamental 
vectors ${\bf e}_1=d(1,0)$ and ${\bf e}_2=d(\frac{1}{2},\frac{\sqrt 3}{2})$. 
The radius of the disks is assumed to be equal to unity and $d$ is the distance 
between their centers. We shall work in the finite horizon regime, 
$2<d<\frac{4}{\sqrt 3}$, for which the diffusion coefficient is finite. 
Some of the disks play the role of catalysts: when the point particle 
collides on one of them, it changes its color instantaneously with a 
probability $p_0$. The catalysts form a regular triangular superlattice 
over the disks lattice, with as fundamental vectors 
${\bf E}_1=nd(0,-\sqrt 3)$ and ${\bf E}_2=nd(\frac{3}{2}, 
\frac{\sqrt 3}{2})$, where $n$ is an integer parameter controlling 
the density of catalysts: in the direction of ${\bf E}_1$ and ${\bf E}_2$, 
one disk over $n$ is a catalyst. Globally, one disk over $N=3n^2$ is 
a catalyst. The configurations with $n=1$ and $n=2$ are depicted 
in Fig.\ \ref{fundamental}.  

Our goal here is to study the cumulative reactive eigenfunctions of this 
system. As done in Ref.\ \cite{Claus1}, we will construct the Frobenius-Perron 
operator governing the time evolution of the color density describing the 
reactive process in this system. We will then perform a spectral analysis 
of this operator. 
%%%%%%%%%%%%%%%%%%%%%%%%%%%%%%%%%%%%%%%%%%%%%%%%%%%%%%%%%%%%%%%%%%%%%%%%%%%%%%%

\subsection{The reactive operator}
The variables needed to describe the point particle in this system are 
its position, its velocity and its color, i.e. $(x,y,v_x,v_y,c)$, where $c$ 
is a discrete variable taking the values $+1$ or $-1$ when the particle color 
is $A$ or $B$. However, the collisions on the disks being elastic,
 the energy of the point particle is conserved. The magnitude of its 
velocity is a constant of motion, that we suppose equal to one, 
$\|{\bf v}\|=1$. The velocity can therefore be described by only one 
coordinate, which can be for example the angle between the velocity and the 
$x$-axis. 
This system is periodic, which means that instead of describing the dynamics 
in the full space, we can restrict our description to an elementary cell of 
the superlattice of catalysts, containing $N$ disks 
(see Fig.\ \ref{fundamental}). 
Moreover, the relevant information concerning the dynamics is contained 
in the collision dynamics: between two collisions, the point particle 
performs a free flight. The three-dimensional flow dynamics can 
therefore be reduced to a Poincar\'e-Birkhoff map, describing the dynamics 
from collision to collision. We will use the Birkhoff coordinates 
${\bf x}=(j,\theta,\varpi)$, $1 \le j \le N=3n^2$, $0 \le \theta < 2\pi$, 
$-1 \le \varpi \le 1$, where $j$ is the index of the disk of the elementary 
cell on which the collision takes place, $\theta $ is an angle giving the 
position of the impact on this disk and $\varpi$ is the sine of the angle 
between the outgoing velocity and the normal at the impact. 
In these coordinates, the mapping of the reactive Lorentz gas is given by 
\begin{equation}
\cases{
{\bf x}_{n+1}=\bphi({\bf x}_{n})\ , \cr
t_{n+1}=t_{n}+T({\bf x}_{n})\ , \cr
{\bf l}_{n+1}={\bf l}_{n}+{\bf a}({\bf x}_{n})\ , \cr
c_{n+1}=\varepsilon({\bf x}_{n})\ c_{n}\ . \cr}
\label{map}
\end{equation}
$t_{n}$ is the time of the $n^{\rm th}$ collision.  $T({\bf x})$ is the 
first-return time  function.  The vector ${\bf l}_{n}$ gives the cell of 
the superlattice visited at time $t_{n}$ and ${\bf a}({\bf x}_n)$ is a 
vector giving the jump carried out on the superlattice during the free 
flight from ${\bf x}_{n}$ to ${\bf x}_{n+1}$.  The change of color is 
controlled by the function:
\begin{equation}
\varepsilon({\bf x}_{n})=\cases{-1& if 
$j_n$ is a catalyst, with a probability $p_0$ , \cr
+1& otherwise .\cr} 
\label{epsilon}
\end{equation}

We notice that the dynamical randomness of the
reactive Lorentz gas is characterized by the following Kolmogorov-Sinai 
entropy per unit time
\begin{equation}
h_{\rm KS} \ = \ \lambda \ + \ \nu \ \left[ p_0 \ln
\frac{1}{p_0}+(1-p_0)
\ln \frac{1}{1-p_0}\right] 
\end{equation}
where $\lambda$ is the unique positive Lyapunov exponent of the 
two-dimensional Lorentz gas and $\nu$ is the frequency of encounters with a 
catalytic disk.

We now want to obtain the Frobenius-Perron operator associated with this map. 
We first start from a description of the particle in the full space. 
The current state and position of the particle are given 
by a suspended flow in terms of the coordinates $({\bf x},\tau,{\bf l},c)$, 
where $0 \le \tau < T({\bf x})$ is the time elapsed since the last collision, 
${\bf l}$ is a vector of the superlattice and $c= \pm 1$. The phase-space 
probability density depending on these variables $q({\bf x},\tau,{\bf l},c)$ 
evolves in time under the Frobenius-Perron operator defined by Eq.\ 
(\ref{FPmatrix}) \cite{Claus1}. 
This continuous-time operator can be reduced to an operator describing 
the dynamics from collision to collision under the mapping (\ref{map}).  
The details are presented in Ref.\ \cite{Claus1}. The operator for the modes 
of wavenumber ${\bf k}$ is given by 
\begin{eqnarray}
\hat{R}_{{\bf k},s}\ \tilde{q}_{{\bf k},s}({\bf x},c)&=&
\exp[-sT(\bphi^{-1}{\bf x}) -i{\bf k}\cdot{\bf a}(\bphi^{-1}{\bf x})]\\
                & &\times \ \{[1-p(\bphi^{-1}{\bf x})] \nonumber 
                    \tilde{q}_{{\bf k},s}(\bphi^{-1}{\bf x},c)
                                 \ + \ p(\bphi^{-1}{\bf x})\ 
\tilde{q}_{{\bf k},s}(\bphi^{-1}{\bf x},-c)\}
\end{eqnarray}
where
\begin{equation}
p({\bf x}=\{j,\theta,\varpi\})=\cases{ p_0& if 
 $j$ is a catalyst,\cr
0& otherwise.\cr}
\end{equation}
The operator $\hat{R}_{k,s}$ is similar to a Frobenius-Perron operator of 
the mapping $\bphi$ up to 
important factors which take into account the varying time of flight between 
the collisions, the spatial modulation by the wavenumber $\bf k$, as well as 
the color change on a part of the domain of definition of the 
Poincar\'e-Birkhoff mapping. If we now consider the total phase-space density 
$f({\bf x})=\tilde{q}({\bf x},A)+\tilde{q}({\bf x},B)$ and the difference of 
phase-space densities $g({\bf x})=\tilde{q}({\bf x},A)-\tilde{q}({\bf x},B)$, 
their time evolutions are decoupled and
we can write an operator governing the  evolution of 
$g({\bf x})$, describing the reactive modes of wavenumber ${\bf k}$ in 
this system \cite{Claus1}
\begin{equation}
\hat{W}_{{\bf k},s}\ g({\bf x})=[1-2p(\bphi^{-1}{\bf x})]\ 
\exp[-sT(\bphi^{-1}{\bf x})
-i{\bf k}\cdot{\bf a}(\bphi^{-1}{\bf x})] \ g(\bphi^{-1}{\bf x}) \ , 
\label{react-op}
\end{equation}
which we call the reactive operator.
%%%%%%%%%%%%%%%%%%%%%%%%%%%%%%%%%%%%%%%%%%%%%%%%%%%%%%%%%%%%%%%%%%%%%%%%%%%%%%
\subsection{Spectral analysis of the reactive operator}
To identify the eigenmode of the reactive operator (\ref{react-op}) which 
controls the 
slowest relaxation of the color, we consider the eigenvalue 
problem for $\hat W_{{\bf k},s_{\bf k}}$ and its adjoint 
$\hat W_{{\bf k},s_{\bf k}}^{\dag}$ imposing $1$ as eigenvalue
\begin{eqnarray}
\hat{W}_{{\bf k},s_{\bf k}}\ \psi_{\bf k}({\bf x})&=&\psi_{\bf k}({\bf x})
\ ,\\
\hat{W}_{{\bf k},s_{\bf k}}^{\dag}\ \tilde{\psi}_{\bf k}({\bf x})
&=&\tilde{\psi}_{\bf k}({\bf x})\ ,
\end{eqnarray}
assuming the normalization condition
\begin{equation}
\langle\tilde{\psi}_{\bf k}^*\psi_{\bf k}\rangle=1 \ .
\end{equation}
A formal solution for $\psi_{\bf k}$ (respectively $\tilde{\psi}_{\bf k}$) 
can be obtained by applying  successively $\hat{W}_{{\bf k},s}$ 
(respectively $\hat{W}_{{\bf k},s}^{\dag}$) to the unit function:
\begin{eqnarray}
\psi_{\bf k}({\bf x})&=&\lim_{n\to\infty}\prod_{m=1}^n[1-2p(\bphi^{-m}
{\bf x})] \ \exp[-s_{\bf k}T(\bphi^{-m}{\bf x})-i{\bf k}\cdot{\bf a}
(\bphi^{-m}{\bf x})] \ , \label{eigenfunction}\\
\tilde{\psi}_{\bf k}({\bf x})&=&\lim_{n\to\infty}\prod_{m=0}^{n-1}
[1-2p(\bphi^m{\bf x})]\ \exp[-s_{\bf k}^*T(\bphi^m{\bf x})+i{\bf k}\cdot
{\bf a} (\bphi^{m}{\bf x})]\ .
\end{eqnarray}
The normalization condition gives an equation to obtain $s_{\bf k}$:
\begin{equation}
1=\lim_{n\to\infty}\langle\prod_{m=-n+1}^{n}[1-2p(\bphi^{-m}{\bf x})]
\ \exp[-s_{\bf k}T(\bphi^{-m}{\bf x})-i{\bf k}\cdot{\bf a}
(\bphi^{-m}{\bf x})]    \rangle \ .\label{norm}
\end{equation} 
We notice that if we consider segments of trajectories of time $t$ from 
initial conditions ${\bf r}_0$ we have that
\begin{eqnarray}
t &\simeq&\sum_{m=-n+1}^{n} T(\bphi^{-m}{\bf x}) \ , \label{corr1} \\
{\bf r}_t-{\bf r}_0 &\simeq&-\; \sum_{m=-n+1}^{n} {\bf a}(\bphi^{-m}
{\bf x}) \ ,
\label{corr2} \\ (1-2p_0)^{N_t} &\simeq& \prod_{m=-n+1}^{n} [1-2p(\bphi^{-m}
{\bf x})]\ , 
\label{corr3} 
\end{eqnarray}
so that Eq.\ (\ref{norm}) can be transformed into Eq.\ (\ref{VanHovereac}) 
for the reaction rate.

We are interested in the smallest relaxation rate $-s_{\bf k}$ which dominates 
at long times $t$. For the reactive case, the dispersion relation is given by 
\begin{equation}
s_{\bf k}=-2\kappa-{\cal D}^{\rm (r)}{\bf k}^2+{\mathcal{O}}({\bf k}^4) \ 
, \label{disp}
\end{equation}
where $\kappa$ is the reaction rate and ${\cal D}^{\rm (r)}$ is the reactive 
diffusion coefficient. Solving Eq.\ (\ref{norm}) allows us to obtain the 
values of these two coefficients. Fig.\ \ref{DrLor} shows the dependence of 
the reactive diffusion coefficient ${\cal D}^{\rm (r)}$ on the reaction 
probability $p_0$, for two different configurations of catalysts, where 
$n=1$ and $n=2$.  
The value of the diffusion coefficient $\cal D$ is indicated for 
comparison. 
We notice the remarkable result that the behavior is similar to the one 
observed for the multibaker model (see Fig\ \ref{DrvsL}).  The explanation is 
the following.  The case $n=1$, with a single catalytic disk in each cell 
where 
the motion is ballistic, is similar to the multibaker map for $L=1$ 
(see Fig.\ \ref{multi}).  In this case, the reactive diffusive coefficient 
rapidly increases with the transition probability $p_0$ in
both the reactive multibaker model with $L=1$ and the reactive Lorentz gas 
with $n=1$. In contrast, the reaction starts to be controled by the diffusion 
as soon as $n\geq 2$ in the reactive Lorentz gas \cite{Claus1} and, similarly, 
for the reactive multibaker with $L\geq 2$.  In these cases, the reactive 
diffusion coefficient has a milder dependence on the transition probability 
$p_0$ and becomes even smaller than the diffusion coefficient for large 
values of $p_0$.
%%%%%%%%%%%%%%%%%%%%%%%%%%%%%%%%%%%%%%%%%%%%%%%%%%%%%%%%%%%%%%%%%%%%%%%%%%%%%
\subsection{Fractal dimension of the cumulative eigenmodes}
The cumulative function of the reactive eigenmode is obtained by 
integration of (\ref{eigenfunction}) over the angle $\theta$ 
between the initial position and the $x$-axis
\begin{equation}
F_{\bf k}(\theta;j,\varpi)=\frac{1}{2 \pi}\int_0^{\theta}d\theta' 
 \lim_{n\to\infty} \prod_{m=1}^n [1-2p(\bphi^{-m}{\bf x}')]\ 
\exp[-s_{\bf k} \ T(\bphi^{-m}{\bf x}')-i{\bf k} \cdot 
{\bf a}(\bphi^{-m}{\bf x}')] \ , \label{cumulLor}
\end{equation}
which is equivalent to the formula (\ref{cumulreac}) according to the 
correspondences (\ref{corr1})-(\ref{corr3}).  

\begin{figure}[ht]
\centerline{\epsfxsize=10 truecm \epsfbox{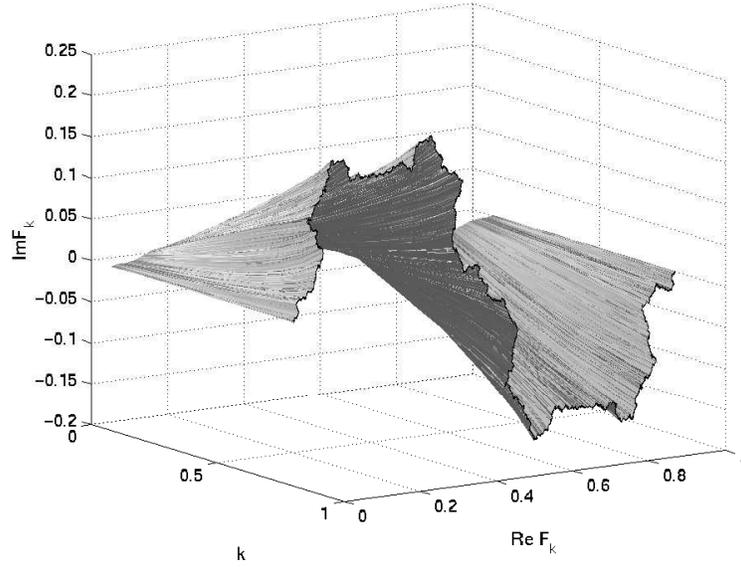}}
\caption{2D reactive periodic Lorentz gas. Three-dimensional representation 
of the cumulative functions of the reactive eigenmodes; 
$({\rm Re}F_k,{\rm Im}F_k)$ are represented in the complex plane as a 
function of the wavenumber $k$, in the case $n=1$, $d=2.3$ and $p_0=0.3$.}
\label{Lorvsk}
\end{figure}

\begin{figure}[ht]
\centerline{\epsfxsize=10 truecm \epsfbox{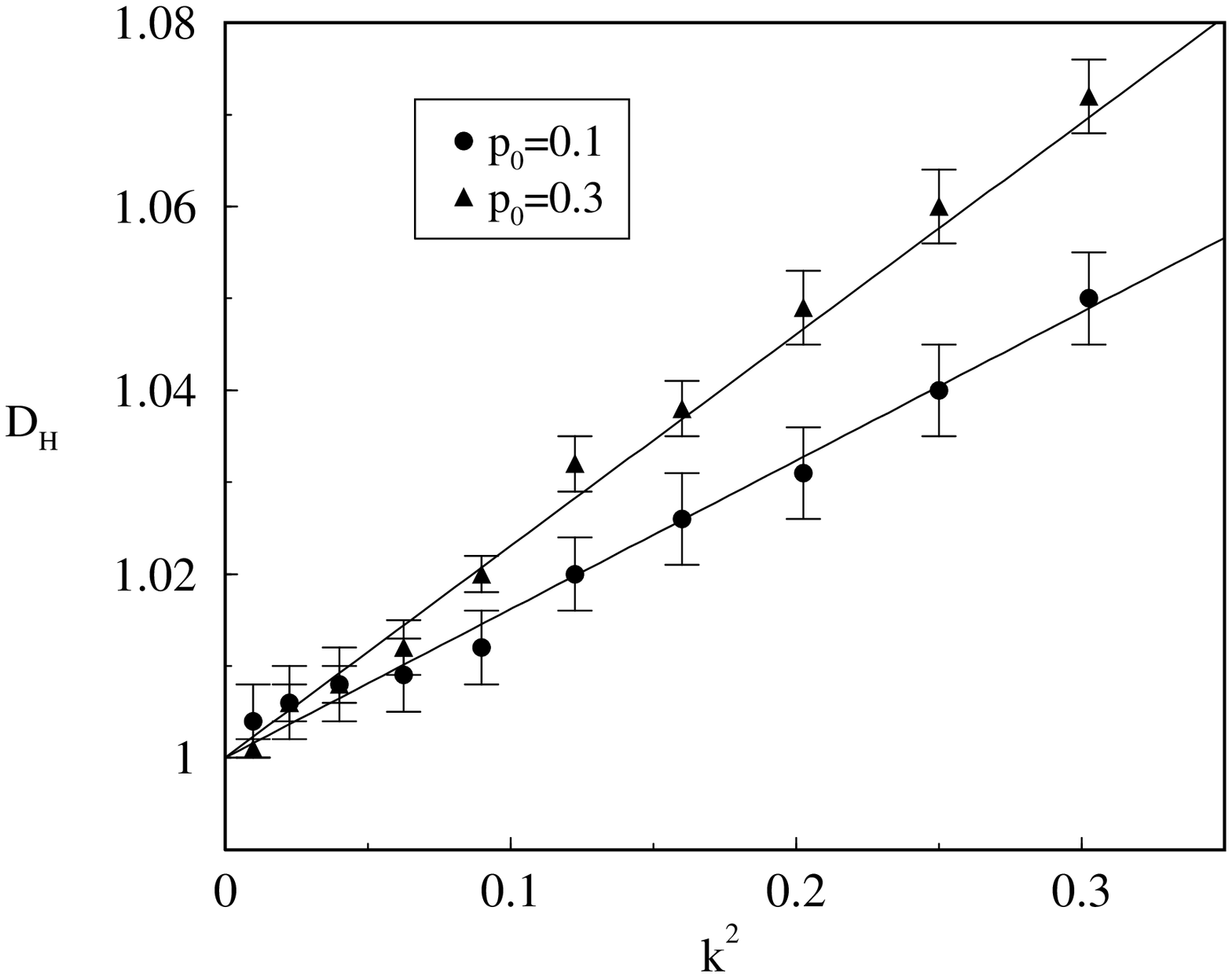}}
\caption{2D reactive periodic Lorentz gas. Dependence of the Hausdorff 
dimension $D_{\rm H}$ of the cumulative functions of the reactive eigenmodes 
on 
the square of the wavenumber $k^2$, for $n=1$, $d=2.3$, $p_0=0.1$ 
and $p_0=0.3$. The symbols are obtained by a box-counting algorithm and 
the solid lines are given by Eq.\ (\ref{hausdim2}) taking into account 
the corrections up to the second order in $p_0$.}
\label{Lorsq}
\end{figure}

Fig.\ \ref{Lorvsk} shows these functions in the complex
plane and as a function of the wavenumber $\bf k$, in the case $n=1$, $d=2.3$ 
and $p_0=0.3$. The Hausdorff dimension of these curves can be computed by a 
box-counting algorithm for different values of the wavenumber ${\bf k}$. 
The values obtained are plotted versus $k^2$ in Fig.\ \ref{Lorsq}, 
in the case $n=1$, $d=2.3$, $p_0=0.1$ and $p_0=0.3$.  We have numerically 
computed the coefficients $\alpha_{\kappa \lambda}$, $\beta_{\kappa \lambda}$ 
and $\sigma^2$, using the expressions 
(\ref{alpha}), (\ref{beta}) and (\ref{sigma2}). 
We have obtained $\alpha_{\kappa \lambda}\simeq 0.23$,  
$\sigma^2 \simeq 0.18$ and $\beta_{\kappa \lambda} \simeq 0.15$. The solid lines 
in Fig.\ \ref{Lorsq} are given by 
Eq.\ (\ref{hausdim2}), considering the corrections up to the second order 
in $p_0$. We observe the remarkable agreement between the theoretical prediction 
and the numerical results. 
%%%%%%%%%%%%%%%%%%%%%%%%%%%%%%%%%%%%%%%%%%%%%%%%%%%%%%%%%%%%%%%%%%%%%%%%%%%%%%%
\section{Conclusions}
\label{conclusions}
In this paper, we have studied in detail simple reaction-diffusion systems in 
which an isomerization is induced by chaotic dynamics.  We have constructed the 
reactive modes of relaxation toward the thermodynamic equilibrium.  These modes 
are given by singular distributions having a cumulative function at small 
values of the wavenumber.  The cumulative functions form fractal curves in 
the complex plane because of the chaotic dynamics and in spite of the 
randomness introduced by the transition probability $p_0$ of
the color in the isomerization.  We explain this remarkable result by the 
fact that the randomness of the system is of a mild type since its 
Kolmogorov-Sinai entropy per unit time (\ref{KS}) remains finite.

We have here derived a formula given by Eq.\ (\ref{hausdim}) for the 
Hausdorff dimension of the reactive relaxation modes in chaotic 
reaction-diffusion systems with two degrees of freedom (\ref{resultreac}). 
We have defined a 
function $Q(\alpha, \beta)$ generalizing the Ruelle topological pressure 
$P(\beta)$. For small values of the wavenumber, we obtain a relationship 
between the Hausdorff dimension $D_{\rm H}$, the reactive diffusion 
coefficient ${\cal D}^{\rm (r)}$, the reaction rate $\kappa$, and two 
derivatives of $Q(\alpha, \beta)$, 
namely $\frac{\partial Q}{\partial \alpha}(1,1)$ 
and $\frac{\partial Q}{\partial \beta}(1,1)$. In the limit of a small reaction 
probability $p_0$, the Hausdorff dimension is related to ${\cal D}^{\rm (r)}$ 
and to the Lyapunov exponent $\lambda$, with corrections in powers of $p_0$. 
For $p_0=0$, we recover the relationship (\ref{diff}) obtained 
in Ref.\ \cite{Gasp1} 
for the diffusive hydrodynamic modes.  We have shown that the reaction rate, 
the reactive diffusion coefficient and the Hausdorff dimension can be 
expressed in terms of quantities which characterize the statistical 
correlations between the number $N_t$ of encounters with
the catalysts, the distance ${\bf r}_t-{\bf r}_0$ travelled in the diffusive 
motion across the system, and the stretching rates $\ln|\Lambda_t|$ 
characterizing the dynamical instability and the chaos. Our formula 
(\ref{hausdim}) thus establishes a new relationship 
between macroscopic reaction-diffusion processes and the microscopic 
dynamical chaos. 

We have tested our results on two reaction-diffusion systems with a simple 
isomerization reaction: a triadic multibaker map and a two-dimensional 
reactive periodic Lorentz gas. In both cases, we study the limit of small 
wavenumbers and small reaction probabilities. The agreement with the theory 
is excellent.

\begin{figure}[ht]
\centerline{\epsfxsize=10 truecm \epsfbox{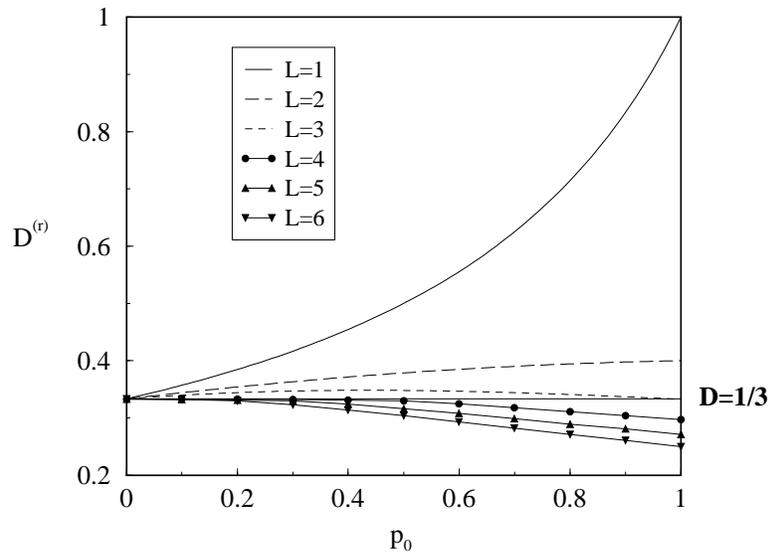}}
\caption{Dependence of the reactive diffusion coefficient ${\cal D}^{\rm (r)}$ 
on $p_0$, for different values of $L$. The value of the diffusion coefficient 
${\cal D}$ is indicated for comparison.} \label{DrvsL}
\end{figure}

\begin{figure}[ht]
\centerline{\epsfxsize=10 truecm \epsfbox{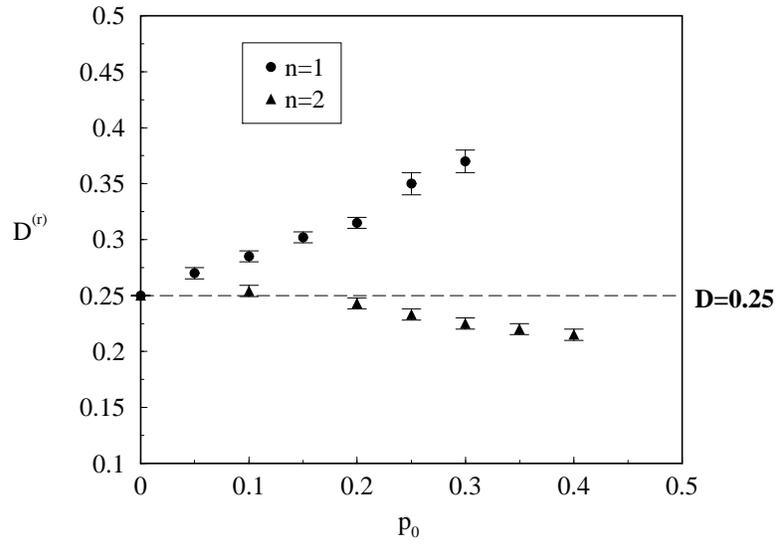}}
\caption{2D reactive periodic Lorentz gas. Dependence of the reactive 
diffusion coefficient ${\cal D}^{\rm (r)}$ on $p_0$, in the case $d=2.3$, 
for $n=1$ and $n=2$. The value of the diffusion coefficient for $d=2.3$ is 
indicated for comparison.}
\label{DrLor}
\end{figure}

Moreover, the study confirms and clarifies results previously obtained for 
similar models of reaction-diffusion \cite{Claus2,Gasp2,Claus1}.  
In particular, our results confirm that the reaction-diffusion process will 
obey on a large spatial scale and a long time scale the macroscopic 
reaction-diffusion equations:
\begin{eqnarray}
\frac{\partial \rho_A}{\partial t}&\simeq& {\cal D}_{AA} \frac{\partial^2 
\rho_A}
{\partial {\bf l}^2}+{\cal D}_{AB} \frac{\partial^2 \rho_B}{\partial 
{\bf l}^2}
-\kappa (\rho_A-\rho_B) \ , \label{EqA}\\
\frac{\partial \rho_B}{\partial t}&\simeq&{\cal D}_{BA} \frac{\partial^2 
\rho_A}
{\partial {\bf l}^2}+{\cal D}_{BB} \frac{\partial^2 \rho_B}{\partial 
{\bf l}^2}
+ \kappa (\rho_A-\rho_B) \ . \label{EqB}
\end{eqnarray}
for the color densities $\rho_A$ and $\rho_B$.  According to  
results (\ref{Drp0}) and (\ref{Delta}), the diffusion coefficients of 
Eqs.\ (\ref{EqB}) are given by
\begin{eqnarray}
{\cal D}_{AA}&=&{\cal D}_{BB}=
\frac{{\cal D}+{\cal D}^{\rm (r)}}{2}= {\cal D}+ \frac{\Delta}{2}\; p_0 + 
{\cal O}(p_0^2)
\\
{\cal D}_{AB}&=&{\cal D}_{BA}=
\frac{{\cal D}-{\cal D}^{\rm (r)}}{2}= - \frac{\Delta}{2}\; p_0 + 
{\cal O}(p_0^2)
\label{crossdiff}
\end{eqnarray}
which explains that the coefficients of cross-diffusion can be small if 
the transition probability $p_0$ is small. Eq.\ (\ref{Delta}) shows that this 
cross-diffusion depends on the correlation between the number $N_t$ of 
catalysts 
encountered and the square $({\bf r}_t-{\bf r}_0)^2$ of the distance 
travelled by the particle in its diffusive motion.  
Moreover,  Figs.\ \ref{DrvsL} and \ref{DrLor} and Eqs.\ (\ref{diffreacp0}) and (\ref{diffreacL}) for the multibaker model show
that the coefficient of cross-diffusion can become significant for 
high transition probability $p_0$ and if the 
density of catalysts is very high, in the case $L=1$ for the multibaker and 
$n=1$ for the Lorentz gas.  We may here say that the 
cross-diffusion, i.e. the coupling between the diffusions of colors $A$ and 
$B$, is induced by the reaction.

In conclusion, the modes of relaxation toward equilibrium turn out to 
display unexpected fractal properties which have their origin in the 
chaotic dynamics of the microscopic motion of particles, even if this motion 
is time-reversal symmetric and volume-preserving.  This result has been 
previously obtained for diffusive processes and, here, we have shown that this 
result extends to reaction-diffusion systems where it allows us to understand 
the relaxation to equilibrium at the level of the microscopic motion.

%%%%%%%%%%%%%%%%%%%%%%%%%%%%%%%%%%%%%%%%%%%%%%%%%%%%%%%%%%%%%%%%%%%%%%%%%%%%%%
\noindent{\bf Acknowledgements.}  We thank Professor G. Nicolis for support 
and encouragement in this research. The authors are financially supported 
by the National Fund for Scientific Research (F.~N.~R.~S. Belgium).  
This research is  supported, in part, by the Interuniversity Attraction 
Pole program of the Belgian Federal Office of Scientific, Technical and 
Cultural Affairs, by the Royal Academy of Belgium, 
and by the F.~N.~R.~S. .
%%%%%%%%%%%%%%%%%%%%%%%%%%%%%%%%%%%%%%%%%%%%%%%%%%%%%%%%%%%%%%%%%%%%%%%%%%%%%%%

%%%%%%%%%%%%%%%%%%%%%%%%%%%%%%%%%%%%%%%%%%%%%%%%%%%%%%%%%%%%%%%%%%%%%%%%%%%%%%%
%FIGURES%
%%%%%%%%%%%%%%%%%%%%%%%%%%%%%%%%%%%%%%%%%%%%%%%%%%%%%%%%%%%%%%%%%%%%%%%%%%%%%%%
\newpage
%Figure 1%
%%%%%%%%%%%%%%%%%%%%%%%%%%%%%%%%%%%%%%%%%%%%%%%%%%%%%%%%%%%%%%%%%%%%%%%%%%%%%%%
%Figure 2%
%%%%%%%%%%%%%%%%%%%%%%%%%%%%%%%%%%%%%%%%%%%%%%%%%%%%%%%%%%%%%%%%%%%%%%%%%%%%%%%
%Figure 3%
%%%%%%%%%%%%%%%%%%%%%%%%%%%%%%%%%%%%%%%%%%%%%%%%%%%%%%%%%%%%%%%%%%%%%%%%%%%%%%%
%Figure 4%
%%%%%%%%%%%%%%%%%%%%%%%%%%%%%%%%%%%%%%%%%%%%%%%%%%%%%%%%%%%%%%%%%%%%%%%%%%%%%%%
%Figure 5%
%%%%%%%%%%%%%%%%%%%%%%%%%%%%%%%%%%%%%%%%%%%%%%%%%%%%%%%%%%%%%%%%%%%%%%%%%%%%%%%
%Figure 6%
%%%%%%%%%%%%%%%%%%%%%%%%%%%%%%%%%%%%%%%%%%%%%%%%%%%%%%%%%%%%%%%%%%%%%%%%%%%%%%%
%Figure 7%
%%%%%%%%%%%%%%%%%%%%%%%%%%%%%%%%%%%%%%%%%%%%%%%%%%%%%%%%%%%%%%%%%%%%%%%%%%%%%%%
%Figure 8%
%%%%%%%%%%%%%%%%%%%%%%%%%%%%%%%%%%%%%%%%%%%%%%%%%%%%%%%%%%%%%%%%%%%%%%%%%%%%%%%
%Figure 9%
%%%%%%%%%%%%%%%%%%%%%%%%%%%%%%%%%%%%%%%%%%%%%%%%%%%%%%%%%%%%%%%%%%%%%%%%%%%%%%%

\begin{thebibliography}{30}
\bibitem{Gilb2}T. Gilbert, J. R. Dorfman and P. Gaspard, Phys. Rev. Lett.
{\bf 85}, 1606 (2000).
\bibitem{Gasp3}P. Gaspard, J. Stat. Phys. {\bf 88}, 1215 (1997).
\bibitem{Gilb1}T. Gilbert, J. R. Dorfman and P. Gaspard, Nonlinearity {\bf 14},
339 (2001).
\bibitem{Gasp1}P. Gaspard, I. Claus, T. Gilbert and J. R. Dorfman,  
Phys. Rev. Lett. {\bf 86}, 1506 (2001).
\bibitem{Gasp4}P. Gaspard and G. Nicolis, Phys. Rev. Lett. {\bf 65}, 
1693 (1990).
\bibitem{Gasp5}P. Gaspard and F. Baras, Phys. Rev. E {\bf 51}, 5332 (1995)
\bibitem{Dorf}J. R. Dorfman and P. Gaspard, Phys. Rev. E {\bf 51}, 28 (1995).
\bibitem{Gasp6}P. Gaspard and J. R. Dorfman, Phys. Rev. E {\bf 52}, 
3525 (1995).
\bibitem{Claus2}I. Claus and P. Gaspard, Phys. Rev. E {\bf 63}, 036227 (2001).
\bibitem{Moran1}B. Moran and W. Hoover, J. Stat. Phys. {\bf 48}, 709 (1987).
\bibitem{Evans1}D. J. Evans, E. G. D. Cohen and G. Morriss, Phys. Rev. A
{\bf 42}, 5990 (1990).
\bibitem{Chernov1}N. I. Chernov, G. L. Eyink, J. L. Lebowitz, and Ya G. Sinai, 
Phys. Rev. Lett. {\bf 70}, 2209 (1993).
\bibitem{Nicolis77} G. Nicolis and I. Prigogine, {\it Self-Organization
in Nonequilibrium Systems} (Wiley, New York, 1977).
\bibitem{Nicolis95} G. Nicolis, {\em Introduction to Nonlinear Science} 
(Cambridge University Press, Cambridge UK, 1995).
\bibitem{ER} J.-P. Eckmann and D. Ruelle, Rev. Mod. Phys. {\bf 57}, 617 (1985).
\bibitem{GaspWang} P. Gaspard and X.-J. Wang, Phys. Rep. {\bf 235}, 291 (1993).
\bibitem{Gaspbook} P. Gaspard, {\it Chaos, Scattering, and 
Statistical Mechanics}(Cambridge University Press, Cambridge UK, 1998).
\bibitem{BarraGasp} F. Barra and P. Gaspard, Phys. Rev. E {\bf 63}, 066215 
(2001).
\bibitem{VanHove} L. Van Hove, Phys. Rev. {\bf 95}, 249 (1954).
\bibitem{Boon} J.-P. Boon and S. Yip, {\em Molecular Hydrodynamics} 
(Dover, New York,1980).
\bibitem{Sinai}Ya. G. Sinai, Russian Math. Surveys {\bf 27}, 21 (1972).
\bibitem{Bowen}R. Bowen and D. Ruelle, Invent. Math. {\bf 29}, 1181 (1975).
\bibitem{Ruelle} D. Ruelle, {\it Thermodynamic Formalism} 
(Addison-Wesley, Reading MA,1978).
\bibitem{Eigen} P. Gaspard, Phys. Rev. E {\bf 53}, 4379 (1996).
\bibitem{Gasp2}P. Gaspard and R. Klages, Chaos {\bf 8}, 409 (1998).
\bibitem{Claus1}I. Claus and P. Gaspard, J. Stat. Phys. {\bf 101}, 161 (2000).
\end{thebibliography}
\end{document}